\documentclass[3p,twocolumn]{elsarticle}

\usepackage{xcolor}
\usepackage{colortbl}
\usepackage{bigstrut}
\usepackage{enumitem}
\usepackage{hyperref}
\usepackage{lipsum}
\usepackage{bm}
\usepackage{multirow}
\usepackage{booktabs}
\usepackage{graphics}

\usepackage{graphicx}
\usepackage{amsmath}
\usepackage{amssymb}

\usepackage{url}

\newcommand{\denseDots}{\ifmmode\mathinner{\ldotp\kern-0.2em\ldotp\kern-0.2em\ldotp}\else.\kern-0.13em.\kern-0.13em.\fi}

\usepackage{framed} 

\usepackage{hyperref}

\usepackage{array}
\newcolumntype{L}[1]{>{\raggedright\let\newline\\\arraybackslash\hspace{0pt}}m{#1}}
\newcolumntype{C}[1]{>{\centering\let\newline\\\arraybackslash\hspace{0pt}}m{#1}}
\newcolumntype{R}[1]{>{\raggedleft\let\newline\\\arraybackslash\hspace{0pt}}m{#1}}

\usepackage[caption=false,font=footnotesize]{subfig}

\journal{arXiv}

\begin{document}

\begin{frontmatter}



\title{A Multi-Scheme Ensemble Using Coopetitive \\ Soft-Gating  With Application to Power Forecasting \\ for Renewable Energy Generation}


\author{Andr\'{e} Gensler}\ead{gensler@uni-kassel.de}
\author{Bernhard Sick\fnref{fn2}}\ead{bsick@uni-kassel.de}

\address{Intelligent Embedded Systems,
University of Kassel, Wilhelmsh\"{o}her Allee 71--73, 34121 Kassel, Germany}
\fntext[fn2]{Intelligent Embedded Systems Homepage: \url{http://www.ies-research.de}}

\begin{abstract}
In this article, we propose a novel ensemble technique with a multi-scheme weighting based on a technique called {coopetitive soft gating}.
This technique combines both, ensemble member competition and cooperation,
in order to maximize the overall forecasting accuracy of the ensemble.
The proposed algorithm combines the ideas of multiple ensemble paradigms (power forecasting model ensemble, weather forecasting model ensemble, and lagged ensemble)
in a hierarchical structure.
The technique is designed to be used in a flexible manner on single and multiple weather forecasting models, and for
a variety of lead times.
We compare the technique to other power forecasting models and ensemble techniques
with a flexible number of weather forecasting models, which can have the same, or varying forecasting horizons.
It is shown that the model is able to outperform those models on a number of publicly available data sets.
The article closes with a discussion of properties of the proposed model which are relevant in its application.
\end{abstract}


\end{frontmatter}


%
%

\begin{figure*}[t]
\begin{minipage}{\textwidth}
\begin{framed}
\section*{Nomenclature}
\scriptsize
\begin{tabular}{lllll}
$t$                  & Time point of evaluation or \emph{forecasting origin}. &  & $D$               & Dimensionality of predictor $\mathbf{x}_{t+k|t}$.       \\
$k$                  & Look-ahead / lead time or \emph{forecasting time step}.       &  & $\psi$               & Idx. of weather forecast. model with $\psi=1,\ldots,\Psi$.              \\
$\mathbf{x}_{t_k|t}$ & NWP for $t_k = t+\Delta k$ made at origin $t$.      &  & $\varphi$         & Idx. of power forecast. model with $\varphi=1,\ldots,\Phi$. \\
$\hat{y}_{t_k|t}$    & Power forecast for $t_k = t+\Delta k$ made at origin $t$.         &  & $j$               & Idx. of ensemble members with $j=1,\ldots,J$.                   \\
$o_{t_k}$                & True observed / measured power value at time $t_k$.                 &  & $w^{(\varphi|\psi)}$ & Weight for model $\varphi$ computed on data of $\psi$.
\end{tabular}
\end{framed}
\end{minipage}
\end{figure*}




\section{Introduction}
\label{sec:intro}


During the past decade, there has been a tremendous growth of the installed capacity of various forms of renewable energy generation. Wind turbines and photovoltaic powerplants contribute substantially to the new mix of energy, which consists of both non-renewable and renewable energy power plants.
Most renewable energy sources have intermittent generation characteristics, i.e., the amount of generated power highly depends on the weather situation and it cannot be regulated the way it is possible with traditional power plants.
In order to guarantee grid stability, the power generation and load in the grid have to be balanced, as the intermediate storage of electrical energy is both inefficient and expensive.
Therefore, highly accurate algorithms have to forecast the available energy on various time horizons. Depending on the forecasting horizon, the forecast is of interest to different actors in the field, e.g., network operators, power plant operators, or electricity traders.
Having an accurate power forecast, the technical and financial risks for all market participants can be reduced.
The power forecasting process typically takes place in two steps:
\begin{enumerate}[leftmargin=*]
\itemsep0em 
\item A meteorological forecast for the desired area (the location of the renewable energy power plant) is computed. This forecast is called \emph{numerical weather prediction} (NWP).
\item The NWP (and optionally other complementary data) is used to forecast the corresponding power generation of the renewable energy power plant using a power forecasting model.
\end{enumerate}
In this article, we focus on the second step of the forecasting process, i.e., we assume the NWP as given.
Naturally, the quality of the forecast typically decreases over time. Depending on the target application, the time horizons are categorized into (very) short-term forecasts in the range up to hours, such as the \emph{intraday forecast}, mid-term forecasts in the range of a few days (including the \emph{day-ahead forecast}), and long-term forecasts in the range of weeks.

Traditionally, the computation of the renewable energy power generation from the NWP is performed using a physical model, i.e., with wind turbine / photovoltaic panel power curves.
While these models yield good performance when the precise parameters for the power plant can be determined, they can easily exhibit systematic errors.
Therefore, machine learning (ML) or statistical approaches became more important in the past decade.
Machine learning models are ``black box'' techniques which train a model based on the historic power generation of the power plant and the respective corresponding NWP forecast.
These trained ML models can then be used to perform a power forecast for future points in time using the NWP forecast for this point in time.
There are a wide variety of models which have exhaustively been
analyzed, e.g., in \citep{Foley2012,Giebel2011, Soman2010}. Typical models are neural networks, multi-linear regressions, or support vector based methods.

Research has shown that the combination of single forecasting models into a so-called forecasting ensemble can improve the forecasting accuracy (e.g., in \citep{Ren2015}).
Traditionally, ensemble techniques are based on diversity principles, such as data diversity, parameter diversity, or structure diversity.
Furthermore, in the area of power forecasting, ensembles are also created using multiple weather forecasting models or lagged ensembles.
Ensemble techniques typically aim at exploiting a particular model principle, namely to
\begin{itemize}[leftmargin=*]
\itemsep0em 
\item aggregate multiple weather forecasts of an ensemble prediction system (EPS), or
\item aggregate multiple weather forecasting model predictions in a multi-model ensemble (MME), or
\item aggregate the forecast for the same forecasting time period from different forecast origins (time-lagged ensemble).
\end{itemize}
In practice, all forecasts may benefit from a combination of those ensemble principles.
As we will show in the following section, the combination of multiple ensemble principles is rarely applied.
Therefore, in this article we propose a novel ensemble technique which combines all the above paradigms using a combination technique which we call \emph{coopetitive soft-gating}.
Coopetition, see e.g., \citep{Loebecke1999} is a term originally emerging from economic research which describes the concept of competitors achieving a joint advantage by cooperating.
We aim to include multiple ensemble paradigms in the way of using the strengths of the global quality of different weather forecasting models and forecasting algorithms, exploiting weather-situation dependent strengths of those models, and making use of
the lead time-dependent properties of each weather and forecasting model. This enables the ensemble to be employed on a variety of forecasting time periods, i.e., day-ahead, or intraday forecasts.

The remainder of this article is structured as follows:
In Section~\ref{sec:sota}, we present the state of the art in forecasting algorithms and ensemble techniques.
Section~\ref{sec:ensembleComp} highlights the way typical ensemble methods differ and how they are computed.
In Section~\ref{sec:mmSoftGating}, we introduce the coopetitive soft gating principle and we show how to apply coopetitive soft gating to create an ensemble.
Section~\ref{sec:result} demonstrates the performance of the proposed approach in comparison to other forecasting and ensemble techniques for intraday and day-ahead forecasting, and using a varying number of weather forecasting models.
Finally, our insights are summarized in Section~\ref{sec:conclusion}. The article closes by giving an outlook on future research directions.

\section{Related Work}
\label{sec:sota}

This section details the state of the art in the area of power forecasting, ensemble methods in general, and ensemble methods for power forecasting.
Some good write-ups on the area of power forecasting are, e.g., given in \citep{Foley2012, Lei2009, Soman2010}.
As each application in power forecasting typically is tied to a certain forecasting horizon, a categorization by the forecasting horizon does make sense.
A good categorization is, e.g., laid out in \citep{Zhang2014}.
Some articles deal with forecasting for particular forecasting horizons, e.g., (very) short-term \citep{Costa2008,Pinson2012,Giebel2011},
mid-term \citep{MIRASGEDIS2006}, and long-term \citep{Craig2002,Hong2014b}.
An alternative categorization is by the methodical category of the power forecasting model.
Possible categories can be, e.g., physical models, such as wind turbine / photovoltaic power curves,
statistical approaches, such as variants of the ARMA model, and machine learning models, such as artificial neural networks, all of which are discussed in the articles mentioned above.

A category of models which are related to machine learning models are ensemble forecasting models.
\emph{Ensemble forecast} is an umbrella term for the aggregation of multiple forecasts to an overall prediction.
Explanatory reasons why ensembles work are founded in bias-variance decomposition \citep{Kohavi1996}, further reasons for the popularity of ensemble methods are also detailed in \citep{Dietterich2000}.
A recent survey gives an overview of the most popular forms of ensembles in the area of classification and regression \citep{Ren2016}.
Ensembles can be formed using a number of principles, the most important ones are:
\begin{itemize}[leftmargin=*]
\itemsep0em
  \item \emph{Data Diversity}: The ensemble is formed by training each ensemble member on a different subset of the training data. Bagging \citep{Breiman1996}, boosting \citep{Schapire1990}, or random forests \citep{Breiman2001} are ensembles of this type.
  \item \emph{Parameter Diversity}: The ensemble is created using different model parameters of the same forecasting model. Multiple Kernel Learning methods \citep{Gonen2011} are part of this diversity technique.
  \item \emph{Structure Diversity}: Different types of forecasting models are used to create the ensemble. These models, sometimes also called heterogeneous ensembles, are detailed, e.g., in \citep{Mendes-Moreira2012}.
\end{itemize}
An overview of ensemble methods for regression is also given in \citep{Mendes-Moreira2012}.

A survey of ensemble techniques in the area of power forecasting is detailed in \citep{Ren2015}.
The literature suggests that ensemble methods can not only yield superior results to single models (e.g., \citep{Taylor2009a,Taylor2002}), ensemble forecasts can also
be used to create probabilistic forecasts to assess uncertainty (e.g., \citep{Pinson2009d,Bessa2012a,Sloughter2010a,Weigel2010}).
This can, for instance, be exploited to estimate the required reserve energy \citep{Matos2011,Holttinen2012}.
In the area of meteorological sciences, the term ensemble typically is used to characterize the form of aggregation of a number of NWPs in an ensemble.
We will highlight the state of the art for the most popular ensemble forms in the following.
More details on the computation of each ensemble type can be found in Section~\ref{sec:ensembleComp}.
All presented forms of ensembles have in common that the combination of the ensemble member forecasts to an overall point forecast is performed
in the power domain, i.e., after applying the power forecasting model.

\emph{Ensemble Prediction Systems} (EPS), sometimes also called \emph{single-model ensemble}s, are created using a systematic variation of the perturbation parameters of the weather forecasting model generating processes, yielding different NWP. The goal of such an EPS is to assess the possible weather outcomes by including an explicit model spread which reflects the stochastic nature of the forecasting task.
It thereby in principle is a data diversity ensemble, which is, however, created through varying the parameters of the NWP generating process in the sense of a parameter diversity ensemble. 
These forms of ensemble forecasts are typically conducted by a weather forecasting model provider. Each of the weather forecasting models is then used for the power generation forecast using a power forecasting model.
An EPS is employed by \cite{Taylor2002} to forecast electrical load using neural networks with multiple scenarios for the weather parameters. The study is performed for a number of lead times up to ten days.
 In \citep{Pinson2009d}, the ensemble forecast is used to predict the forecasting skill using by evaluating the spread of the EPS, the EPS is also compared to lagged ensembles (explanation see below).
\cite{Alessandrini2013} conduct a comparative study between two ensemble prediction systems regarding their forecasting accuracy for wind power.
In \citep{Troldborg2014}, an EPS is used to create probabilistic forecasts to investigate extreme weather situations and ramp events.

\emph{Multi-Model ensemble}s (MME), sometimes also called ``poor-man's ensembles'', refers to the combination of (typically deterministic) point forecasts of different weather forecasting model providers. In principle, each of the single forecasts is created independently. MMEs have characteristics different from EPS, as each ensemble member yields the most likely point forecast and does not try to explicitly include a model spread.
MMEs thus have data diversity characteristics.
Furthermore, multi-model ensemble members can differ in their structure (e.g., different number of NWP parameters, etc.).
     \cite{Pierro2016} use MME for photovoltaic forecasting, the creation of prediction intervals for uncertainty assessment is also investigated.
    An MME of four climate forecast systems using coupled ocean-atmosphere models is investigated in \citep{VanOldenborgh2012} with particular focus on forecast verification.
    The performance of MMEs is compared to EPS forecasts in \citep{Ziehmann2000}.

\emph{Power forecasting model ensemble}s (PME) make the basic assumption that a single forecasting model cannot be an optimal estimator for a given data set due to a too simple model structure, varying intrinsic uncertainties depending on the data set,
or poor generalization characteristics of certain models.
This form of ensemble then uses the predictions of multiple (independently trained) power forecasting algorithms, typically given a single weather forecast, to combine to an overall forecast in the form of a data, parameter, or structure diversity ensemble.
In \citep{Siwek2009}, ensembles of neural networks are used for load forecasting using a number of different combination methods including principal component based methods.
Bayesian adaptive model combination investigating a number of different neural network types is performed in \citep{Li2011} including a unified approach for model selection.
An ensemble forecasting technique with adaptive weighting of the
ensemble members using a locality assessment of the weather situation is investigated on a variety of models in \citep{Gensler2016ensemble},
a variant for probabilistic forecasts is presented in~\cite{genslerpcsge17}.
``Standard'' ensemble techniques, such as bagging \citep{Breiman1996} or boosting \citep{Schapire1990}, also fall under the category of PME. These techniques can also easily be combined with other ensemble techniques.

\emph{Time-lagged ensemble}s (TLE) use a repetitive forecast of the same absolute point in time computed from different forecasting origins to aggregate to an ensemble.
TLE can be computed using only a single power forecasting model and a single weather forecasting model in the form of a data diversity ensemble. 
     \cite{Mittermaier2007} uses lagged ensembles for the aggregation of high-resolution forecasts to achieve the effect of spatial averaging.
    Lagged ensembles are also frequently used to assess the uncertainty of a forecast, e.g., using risk-indices \citep{Pinson2004,Pinson2009d}.
Table~\ref{tab:ensembletypes} summarizes the possible ensemble types by the number of weather and power forecasting models involved.

So-called \emph{Analog Ensembles} \citep{DelleMonache2013,Gensler2016Analog} have a similarity in naming, but do not fall into the same methodical category, as they are related to nearest neighbor techniques.
Ensemble forecasts can in many cases be regarded as post-processing techniques, which means that they aggregate single forecasts of ensemble members to an overall forecast.
In addition to a more precise determination of a forecast, the assessment of the forecasting uncertainty is easier using ensemble techniques, e.g., using Bayesian model averaging \citep{Duan2007}.

\begin{table}[t!]
\centering
\scriptsize
\caption{Categorization of ensemble types by number of weather and power forecasting model.
}
\label{tab:ensembletypes}
\vspace{1pt}
\begin{tabular}{l|rr}
\multicolumn{1}{c|}{Ensemble} & \multirow{2}{*}{ \begin{tabular}{@{}c@{}}Weather \\ Forecasting Models\end{tabular}} & \multirow{2}{*}{\begin{tabular}{@{}c@{}} Power \\ Forecasting Models\end{tabular}}\\
\multicolumn{1}{c|}{Model Type}      &                  &                 \\
       \midrule
No Ens. & $1$                 & $1$                 \\
EPS    & $>1$                 & $\geq 1$                 \\
MME    & $>1$                 & $\geq 1$                 \\
PME    & $1$                 & $>1$                 \\
TLE    & $1$                 & $1$
\end{tabular}
\end{table} 

\section{Ensemble Computation Methods for Power Forecasting}
\label{sec:ensembleComp}

A forecast is performed at forecasting time $t$ which is also called the \emph{forecasting origin}.
The forecast is normally conducted for a number of forecasting time steps (or lead times) with fixed time increment $\Delta$ and

\vspace{.3cm}

\noindent\begin{tabular}{l L{.7cm} @{}L{.8cm} @{}L{.8cm} }
minimum lead time              & $t~+~$ & $k_{min}$ & $\cdot~\Delta$, \\
currently considered lead time & $t~+~$ & $k$ & $\cdot~\Delta$, \\
maximum lead time              & $t~+~$ & $k_{max}$ & $\cdot~\Delta$,
\end{tabular}
\smallskip

\noindent where the lead times $k$ typically are a set with $k \in \{k_{min},\ldots,k_{max} \} \in \mathbb{N}$, where $k_{min},k_{max} \in \mathbb{N}$ and $k_{min} \leq k_{max}$.
We will abbreviate the currently considered lead time as $t_k = t+k \cdot \Delta$.
The minimum lead time $k_{\text{min}}$ and the forecasting horizon $k_{\text{max}}$ can be chosen depending on the application.
An overview of the most important variables used throughout this article is also given in the nomenclature section at the beginning of the article.
The process of creating a forecast is typically conducted using a power forecasting model which transforms the input data of a predictor $\mathbf{x}_{t_k|t} \in \mathbb{R}^D$ ($D$-dimensional input data $\mathbf{x}$ for forecasting time $t_k$ created at forecasting origin $t$)
to a forecast of a target predictand $\hat{y}_{t_k|t}$ in the form
\begin{equation}
\hat{y}_{t_k|t} = f(\mathbf{x}_{t_k|t}|\mathbf{\Theta}),
\end{equation}
where $\mathbf{\Theta}$ describes the set of model parameters of model $f$.
In case the power forecasting model is an ensemble, the deterministic forecast is in many cases created using a weighted sum of single forecasts in a post-processing step as
\begin{equation}
\label{eq:ensembleGeneral}
\bar{\hat{y}}_{t_k|t} = \sum_{j=1}^J w^{(j)} \hat{y}^{(j)}_{t_k|t},
\end{equation}
where $J$ is the number of ensemble members, $w^{(j)}$ is the respective weighting coefficient of the power forecasting model, and
$\hat{y}^{(j)}_{t_k|t}$ is the forecast of the $j$-th ensemble member.
For all types of ensembles, we require that the sum of weights has to fulfill
\begin{equation}
\label{eq:sumEqualsOne}
  \sum_{j=1}^{J} w^{(j)} = 1, ~~~w^{(j)} \in \mathbb{R}^+_0
\end{equation}
in order to not over- or underestimate the forecast on average.
Ensembles create the final prediction by aggregating the forecasts using Eq.~(\ref{eq:ensembleGeneral}).
There are two basic approaches of setting weights:
\begin{itemize}
  \item \emph{Cooperation / Weighting}: One possibility to create an ensemble forecast is by letting the single ensemble members cooperate in creating the final point estimate. In the easiest case,
  the weights $w^{(j)} \in [0,1]$ can be chosen equally, i.e., $w^{(j)} = \frac{1}{J}$. Other possibilities are, e.g., to set them proportional to their overall average forecasting quality, if known.
  The weight values typically are static in this technique, they do not change after being set.
  \item \emph{Competition / Gating}: In this approach, in each situation one model succeeds in competing against the other models, i.e., $w^{(j)}\in \{0,1\}$ and $\sum_{j=1}^J w^{(j)} = 1$.
  The challenge of this approach consequently is in deciding which power forecasting model should win the competition for a particular forecast.
  The weight values are dynamic in this technique, they vary depending on some defined criterion.
\end{itemize}

\begin{figure}[tb]
\centering
\includegraphics[width=.45\textwidth]{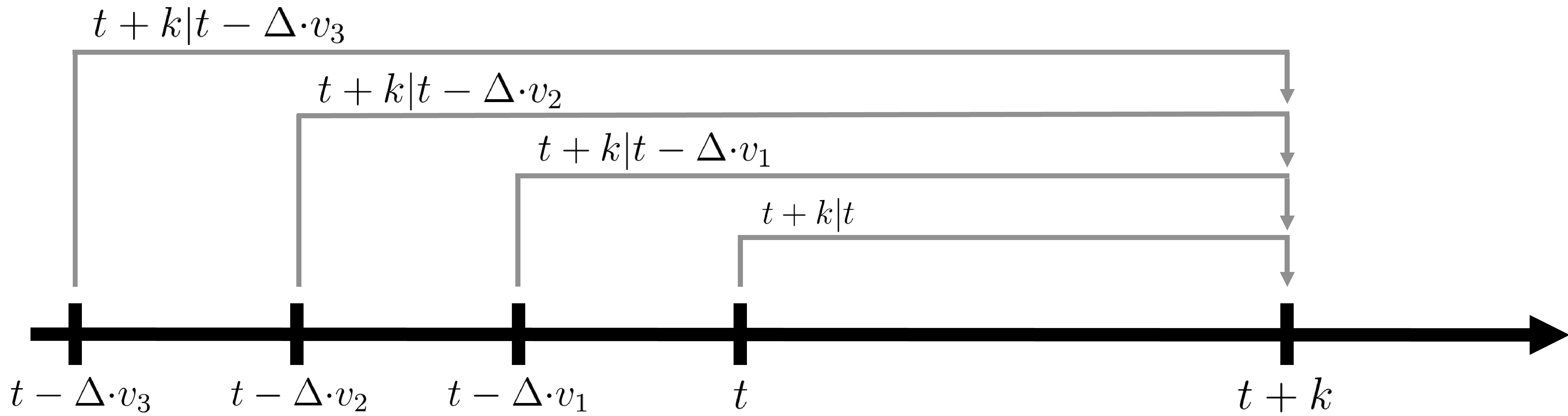}
\vspace{-6pt}
\caption{Aggregation principle of time-lagged ensembles.}
\label{fig:tlevis}
\end{figure}

In the area of power forecasting, the predictor $\mathbf{x}_{t_k|t}$ typically is an NWP forecast, the
predictand is the expected power generation $\hat{y}_{t_k|t}$.
Typical applied ranges in the area of power forecasting are the so-called day-ahead forecast (where $k_{\text{min}}=24$, $k_{\text{max}}=48$, $\Delta = 1 h$),
or the intraday forecast (with $k_{\text{min}}=1$, $k_{\text{max}}=24$, $\Delta = 1 h$).
For some operational day-ahead forecasts, this time period may vary. 
The forecast $\hat{y}^{(j)}_{t_k|t}$ of each ensemble member can be computed in a number of ways depending on the form of the ensemble:
\begin{enumerate}[leftmargin=*]
\itemsep0em 
  \item In case an ensemble prediction system (EPS) is used, a single power forecast is created with
\begin{equation}
\hat{y}^{(j)}_{t_k|t} = f(\mathbf{x}_{t_k|t}^{(j)}|\bm{\Theta}).
\end{equation}
For this type of ensemble, the input NWP values are changing, however, the type and parametrization of the power forecasting model normally remains the same.
As for many EPS forecasts all NWP ensemble members are assumed to have an equal
probability of being correct, the values of $w^{(j)}$ are often chosen equally, i.e., $w^{(j)} = \frac{1}{J}$, if a deterministic forecast is desired.
  \item For Multi-Model Ensembles (MME), a single power forecast is created using
\begin{equation}
\hat{y}^{(j)}_{t_k|t} = f_j(\mathbf{x}_{t_k|t}^{(j)}|\bm{\Theta}^{(j)}).
\end{equation}
For MMEs, the NWPs may be of a different form (and / or they may have different number of dimensions). The power forecasting models consequently have a different structure, thus, different model
parameters $\bm{\Theta}^{(j)}$ have to be chosen for each ensemble member. This does not necessarily have to be the case for an EPS.
MME members often have a varying overall quality due to a varying weather forecasting model quality. The corresponding weights $w^{(j)}$ therefore typically have different values which can be set
according to the expected quality of the models (e.g., by testing the model on some historic time period) in order to maximize the ensemble quality.
\item Forecasting model ensembles (PME) typically operate on the same input data in the form
\begin{equation}
\hat{y}^{(j)}_{t_k|t} = f_j(\mathbf{x}^{(j)}_{t_k|t}|\bm{\Theta}{^{(j)}}).
\end{equation}
This form of ensemble can be computed on a single NWP (then, all $\mathbf{x}^{(j)}_{t_k|t}$ are equal), or on a subset of the NWP parameters (in subspace methods \citep{Ho1998}, for instance).
In this case, the dimensionality of the NWP is $\mathbf{x}^{(j)}_{t_k|t} \in \mathbb{R}^{D'}$ with $D' < D$.
The main differentiating factor of the ensemble members is their form of data, parameter, or structure diversity (see Section~\ref{sec:sota}).
\item Finally, (time-)lagged ensembles (TLE) typically operate on the same NWP using the same power forecasting model. They use forecasts for the forecasting time $t_k$ from different forecasting origins $t-\Delta \cdot \mathbf{\upsilon}_j$ in the form
\begin{equation}
\hat{y}^{(j)}_{t_k|t} = f(\mathbf{x}_{t_k|t-\Delta \cdot \mathbf{\upsilon}_j}|\bm{\Theta}),
\end{equation}
such as shown in Fig.~\ref{fig:tlevis}.
The value of $\mathbf{\upsilon}_j \in \mathbb{N}^+_0$ denotes the amount of lag of the $j$-th ensemble member. Typically, the corresponding weights $w^{(j)}$ are chosen in a form that smaller values of  $\mathbf{\upsilon}_j$ have a higher weight
(as the amount of time-lag is smaller for those forecasts, which typically correlates with an increased precision of the forecast).
\end{enumerate}

It is being shown that each type of ensemble aims at exploiting a different aspect of the present data.
As can be seen, the two basic approaches weighting and gating both have distinct advantages, however they are either not dynamic (weighting), or do not allow for cooperation (gating).
In the following, we present a weighting scheme which combines the advantages of both cooperation and competition in the form of a \emph{coopetitive soft gating} technique.
This scheme is applied in an ensemble structure that combines multiple of the ensemble principles laid out above.
\begin{figure*}[htb]
\centering
\begin{minipage}{.49\textwidth}
\centering
\includegraphics[width=\textwidth]{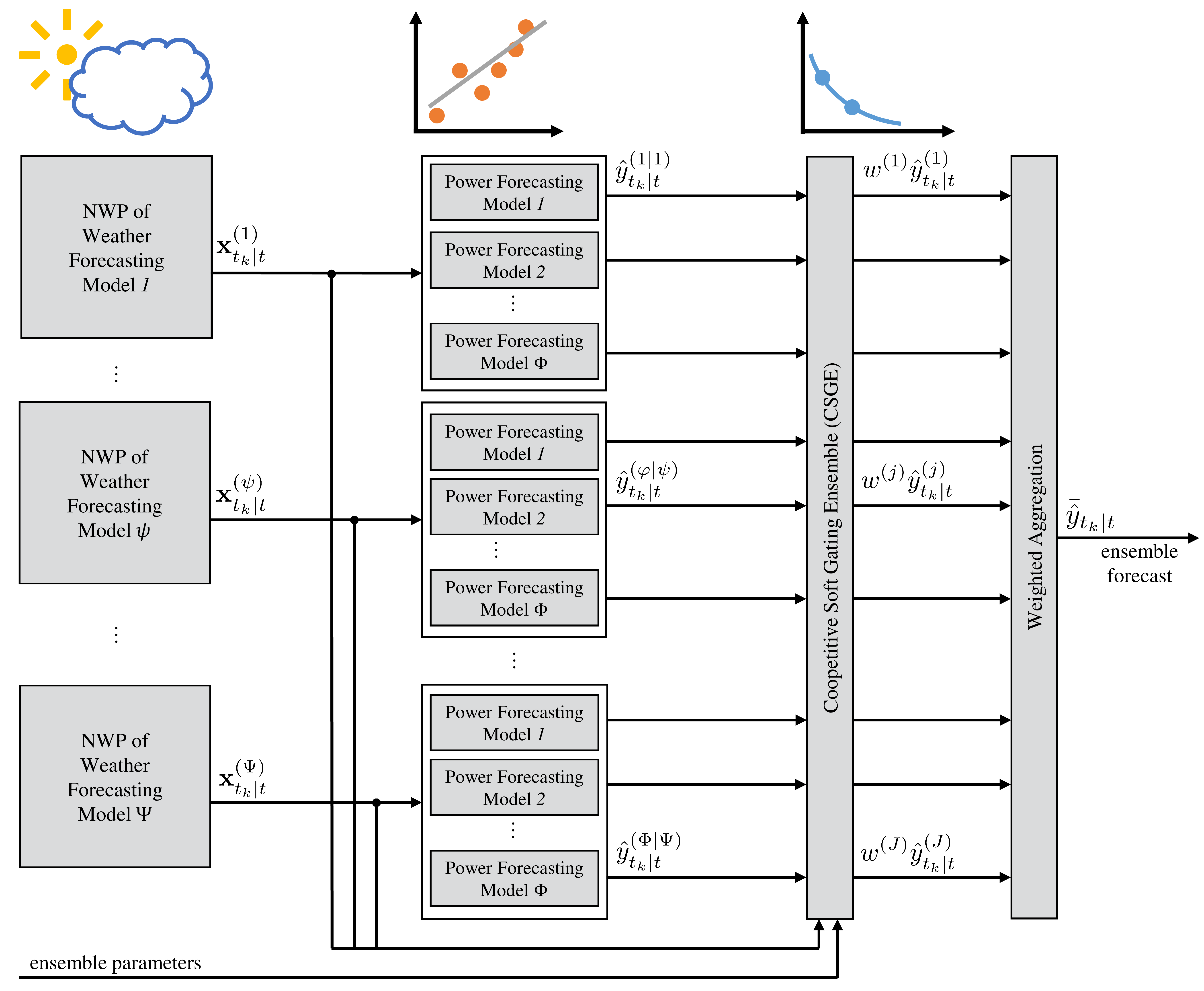}
\caption{Overview of the proposed ensemble model. As can be seen from the figure, the ensemble has a hierarchical structure, where a number of predictors from $\Psi$ weather forecasting models (EPS, MME, or TLE) are used
to forecast a common target predictand. Therein, $\Phi$ power forecasting models
each create a power forecast for a particular weather forecast. The total number of $J = \Psi \cdot \Phi$ forecasts are combined in a post-processing stage using
coopetitive soft gating.
 The ensemble member weights are computed as described in Section~\ref{sec:mmSoftGating}.}
\label{fig:mmescheme}
\end{minipage}
\hspace{.5cm}
\begin{minipage}{.45\textwidth}
\centering
\includegraphics[width=.93\textwidth]{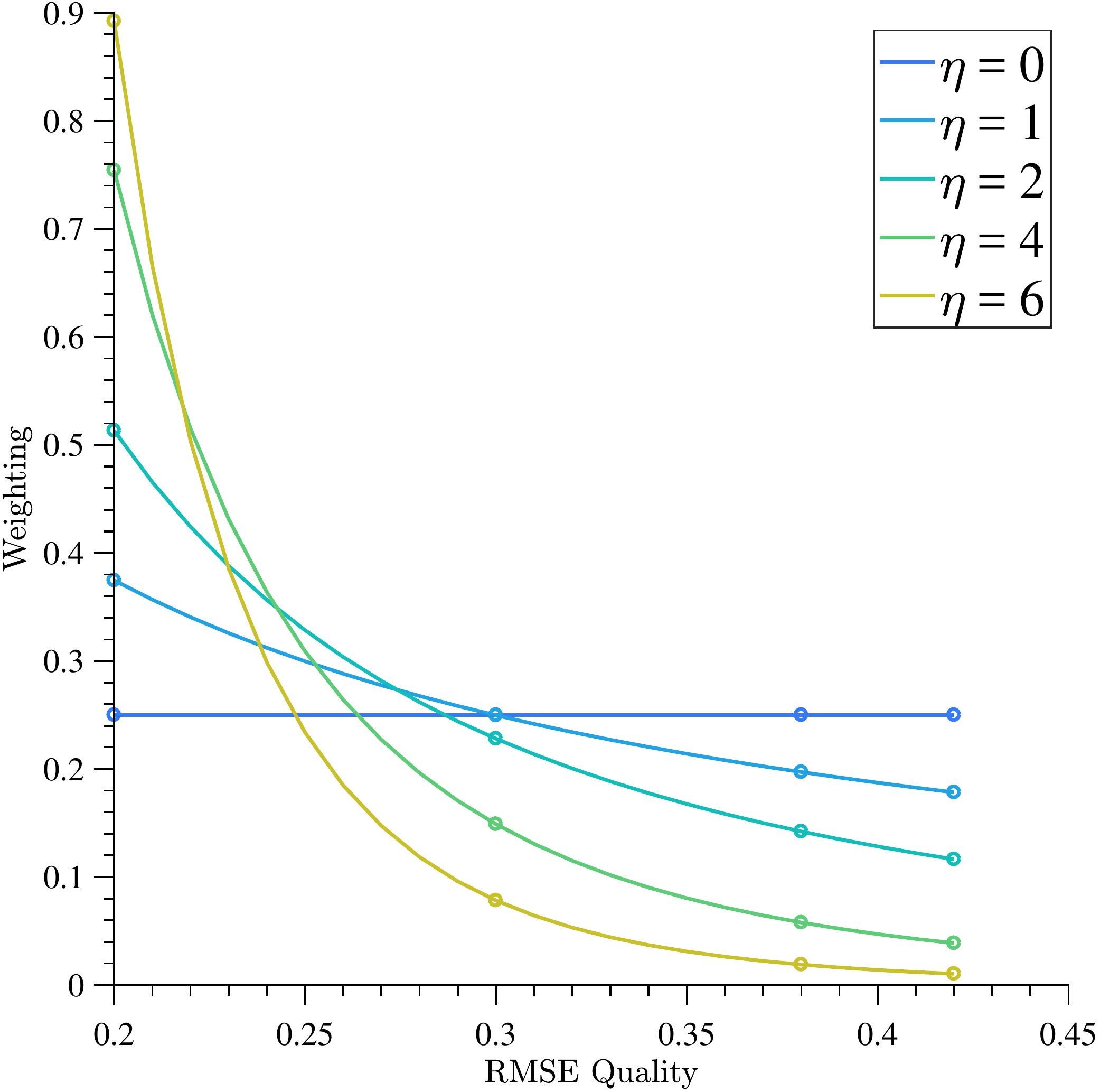}
\caption{Weighting example of the parameter $\eta$, which is used in the coopetitive soft gating formula (Eq.~(\ref{eq:weightingFormulaGeneral})).
The graphs represent the values of the weighting formula depending on the value of $\eta$.
The different ensemble members compete for influence in the ensemble.
The higher the value of $\eta$, the higher the weighting of well performing models.
}
\label{fig:weightingETA}%
\end{minipage}
\end{figure*}

\section{Coopetitive Soft Gating Ensemble}
\label{sec:mmSoftGating}

As outlined in Section~\ref{sec:ensembleComp}, ensemble techniques typically aim to exploit a single principle for ensemble generation.
The proposed technique aims at using a multitude of weighting principles.
The principal structure of the proposed ensemble technique is visualized in Fig.~\ref{fig:mmescheme}.
The weighting of the ensemble remains the same as in Eq.~(\ref{eq:ensembleGeneral}).
However, we have a hierarchical ensemble structure: For each weather forecasting model $\psi=1,\ldots,\Psi$ (which can be an arbitrary NWP of an EPS, MME, or TLE, e.g., of an intraday or day-ahead model, for a particular time step to be forecasted),
a number of power forecasting models $\varphi=1,\ldots,\Phi$ are used to forecast the target predictand for each weather forecasting model. The power forecasting models do not necessarily have to be
the same for each weather forecasting model, but, for the sake of easier understanding, we will use the same type and number of power forecasting models for each weather forecasting model here.
The overall number of ensemble participants $J$ consequently is $J = \Psi \cdot \Phi$.
The individual predictions of each power forecasting model are then aggregated and fused to an overall forecast in a post-processing step according to Eq.~(\ref{eq:ensembleGeneral}).
The main innovation here is the way the single weights $w^{(j)}$ are constructed.
In order to clarify the origin of each weighting term with respect to the weather forecasting model $\psi$ and the power forecasting model $\varphi$, we define the weight of an ensemble member as $w^{(j)} = w^{(\psi,\varphi)}$.
The idea is as follows: For each ensemble participant we construct the coopetitive soft gating (explained in Section~\ref{sec:coopSoftGating}) considering the following three aspects for both, power forecasting models \emph{and} weather forecasting models, respectively (leading to $6$ aspects):
\begin{enumerate}[leftmargin=*]
\itemsep0em 
  \item \emph{Global Weighting}: The ensemble weights are determined for the respective model regarding the overall observed performance of a model during ensemble training. This is a fixed weighting term. Thereby, overall strong models have more influence than weaker models. This form of weighting is described in Section~\ref{sec:glob}.
  \item \emph{Local Soft Gating}: The ensemble members are weighted depending on the model input (the NWP forecast) $\mathbf{x}^{(j)}_{t_k|t}$.
  This form of weighting assesses the quality of a model considering the current input, i.e., a local quality assessment is performed.
  The idea is that a number of models may have different strengths in a different set of weather situations (e.g., due to ensemble diversity effects). This form of weighting is described in Section~\ref{sec:loc}.
  \item \emph{Lead time-dependent Soft Gating:} Models may have a lead time-dependent quality development.
  The goal of this form of soft gating is to weigh the model depending on the lead time~$k$.
  In the case of power forecasting models, methods such as the persistence method perform very well on short time horizons, while they quickly loose their quality for longer time-horizons.
  Additionally, weather forecasting models such as intraday models typically perform very strong on short time horizons due to very recent weather measurements.
  This form of weighting is described in Section~\ref{sec:dt}.
\end{enumerate}

The overall weighting term for each ensemble member can be described as
\begin{equation}
\label{eq:overallweighting}
w^{(j)} = w^{(\psi,\varphi)} = \frac{w^{(\psi)} \cdot w^{(\varphi|\psi)}}{\sum_{\psi^*=1}^{\Psi} \sum_{\varphi^*=1}^{\Phi}  w^{(\psi^*)} \cdot w^{(\varphi^*|\psi^*)}},
\end{equation}
where $w^{(j)} = w^{(\psi,\varphi)}$ is the overall weight for ensemble member $j$ computed using power forecasting model $\varphi$ and weather forecasting model $\psi$. The weights $w^{(\psi)}$ are the
\emph{weather forecasting model} dependent weights, $w^{(\varphi|\psi)}$ are the \emph{power forecasting model} dependent weighting factors of power forecasting model $\varphi$ computed on weather forecasting model $\psi$.
The denominator is a normalization term which ensures $\sum_{j=1}^{J} w^{(j)} = 1$, see Eq.~(\ref{eq:sumEqualsOne}).
The weights of the weather forecasting model can be decomposed into
\begin{equation}
  w^{(\psi)} = w^{(\psi)}_g \cdot w^{(\psi)}_l \cdot w^{(\psi)}_{ k},
\end{equation}
while the weights of the power forecasting model are computed with
\begin{equation}
  w^{(\varphi|\psi)} = w^{(\varphi|\psi)}_g \cdot w^{(\varphi|\psi)}_l \cdot w^{(\varphi|\psi)}_{ k}.
\end{equation}
The indices $g$, $l$, $k$ denote the respective weighting aspects global weighting ${g}$ (Section~\ref{sec:glob}), local soft gating ${l}$ (Section~\ref{sec:loc}), or lead time-dependent soft gating ${ k}$ (Section~\ref{sec:dt})
for both, weather forecasting model $\psi$ and power forecasting model $\varphi$.
Using multiple weather forecasting models and multiple power forecasting models, the overall number of weights per weather forecasting model and power forecasting model combination consequently adds up to six.
The ensemble training process is described in Section~\ref{sec:overallmdl}.
In addition to the description of the weighting factors in the different sections, Fig.~\ref{fig:intradayfig} illustrates an example of the functionality of the different weighting aspects and the overall proposed technique.
Fig.~\ref{fig:smGlob} shows the development of the global weighting aspect,
Fig.~\ref{fig:smLoc} shows the development of the local soft gating, and Fig.~\ref{fig:smDt} illustrates the lead time-dependent weighting aspects.
Section~\ref{sec:coopExample} gives an overall application example of the proposed coopetitive soft gating ensemble (CSGE) technique.

\subsection{Coopetitive Soft Gating Principle}
\label{sec:coopSoftGating}
The goal of coopetitive soft gating is to weigh ensemble members according to their performance in a way that
combines the properties of both principles, weighting and gating.
The weighting technique should allow for a flexible application to all of the three weighting principles (global, local, and lead time-dependent) proposed in the previous section.
The weights are built from a quality estimate of the different ensemble members, which typically is represented in the form of an error, e.g., the root mean square error (RMSE).
Having a quality estimate for each ensemble member, the proposed weighting technique should fulfill the following requirements:
\begin{enumerate}[leftmargin=*]
\itemsep0em 
	\item It must return a score ordered from high weights (low error) to low weights (high error), i.e., the inverse of what the error score is initially represented in.
	\item It must be able to weigh errors nonlinearly, as the optimal weighting possibly can not be represented in a linear relationship. The amount of nonlinear weighting should be controllable by the user.
	\item It must be insensitive to the value range of the error scores, i.e., it should only factor in the relative quality differences of the contributing ensemble members.
	\item It must retain the value range of the ensemble prediction, i.e., it has to fulfill Eq.~(\ref{eq:sumEqualsOne}).
\end{enumerate}
We fulfill criteria 1. -- 3. using the weighting function
\begin{equation}
\label{eq:coopPre}
	\varsigma'_{\eta}(\bm{\Omega},\omega) = \frac{\sum_{{n}=1}^N \bm{\Omega}_n}{\omega^\eta +\epsilon}~,~~~ \eta \in \mathbb{R}^+_0,
\end{equation}
where $\bm{\Omega} \in \mathbb{R}^N$ is a tuple of values containing all $N$ reference quality estimates, and $\omega$ is an arbitrarily evaluated point of the function. Typically, $\omega$ can be chosen to be an element of $\bm{\Omega}$.
The real number $\eta \geq 0$ represents the amount of exponential weighting. The higher the value of $\eta$, the higher the weight of models models with low errors.
$\epsilon$ is a very small number that avoids division by $0$ in the unlikely case that $\omega$ is $0$.
Assuming that $\omega \in \bm{\Omega}$ and that $\varsigma'_{\eta}(\bm{\Omega},\omega)$  is computed for each element in $\bm{\Omega}$,
we achieve criterion 4.~by adjusting the weighting function by normalizing with the sum of the weights in $\bm{\Omega}$, i.e.,
\begin{equation}
  \varsigma_\eta(\bm{\Omega},\omega) = \frac{\varsigma'_{\eta}(\bm{\Omega},\omega)}{\sum_{n=1}^N \varsigma'_{\eta}(\bm{\Omega},\bm{\Omega}_n)}.
\end{equation}
This form of the coopetitive soft gating formula
can further be simplified to
\begin{equation}
\label{eq:weightingFormulaGeneral}
	\varsigma_\eta(\bm{\Omega},\omega) = \frac{1}{(\omega^\eta+\epsilon) \sum_{{n}=1}^N ((\bm{\Omega}_{n})^\eta+\epsilon)^{-1}}.
\end{equation}
An example of how Eq.~(\ref{eq:weightingFormulaGeneral}) works depending on the way $\eta$ is set is shown in Fig.~\ref{fig:weightingETA}.
The example uses exemplary quality estimates in the range of $0.2 - 0.42$ which represent the quality of different ensemble members. As can be seen in Fig.~\ref{fig:weightingETA}, 
when increasing $\eta$, the weights of the well-performing ensemble members do increase.
An advantage of this form of weighting is that it only has a single weighting parameter $\eta$ for optimization (unlike methods derived from exponential functions in the form $f(x) = e^{Ax+B}$ , e.g., in \citep{Platt1999}).
The following sections use the coopetitive soft gating formula of Eq.~(\ref{eq:weightingFormulaGeneral}) to compute the ensemble weights $w^{(j)}$ for each of the three weighting aspects (global, local, and lead time-dependent).

\begin{figure*}[tb]
	\centering
	\subfloat[Power forecasting model 1.]{%
	\includegraphics[width=0.235\textwidth]{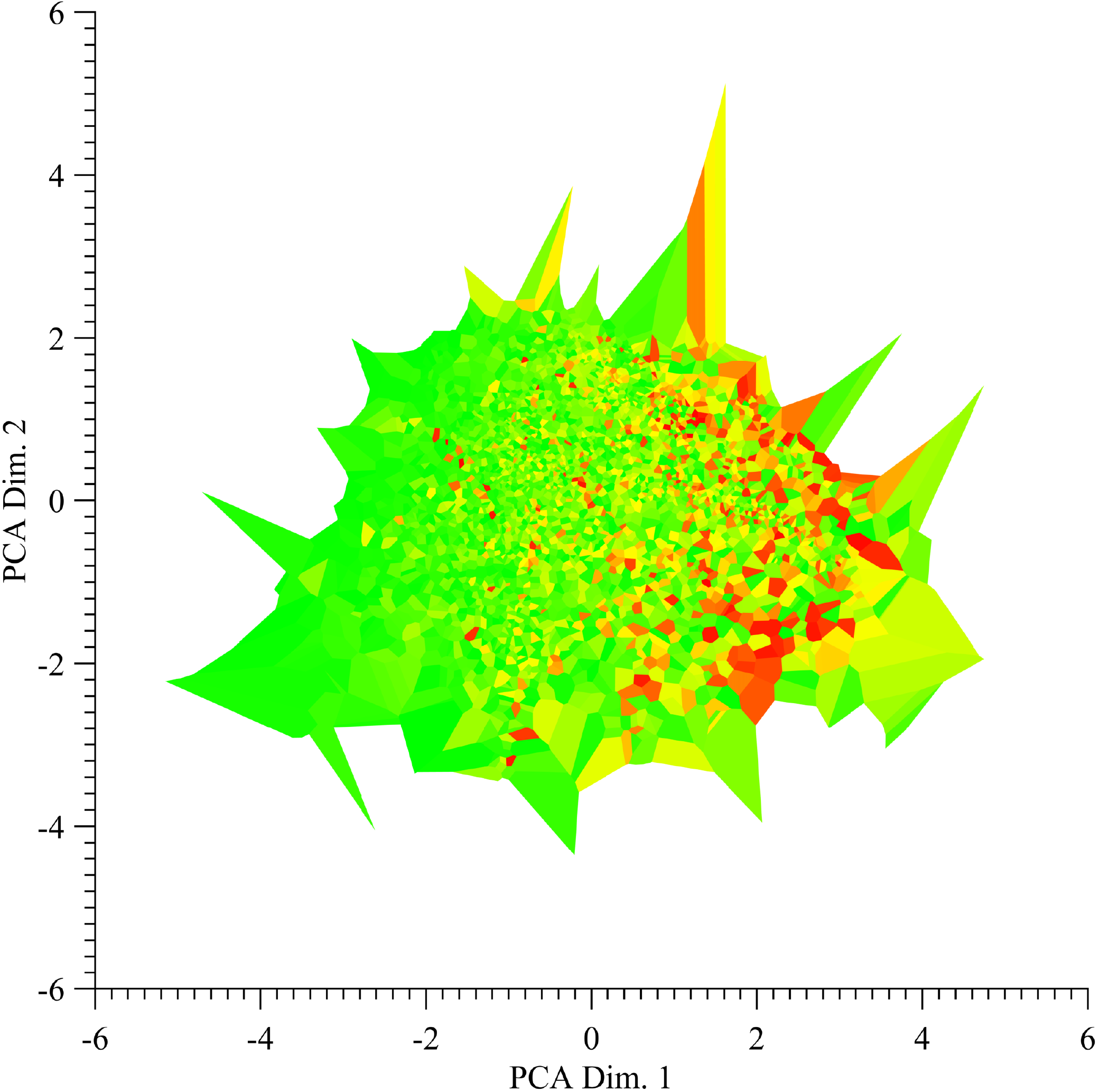}
	\label{fig:rmseANN}%
	}
	\hfil
	\subfloat[Power forecasting model 2.]{%
	\includegraphics[width=0.235\textwidth]{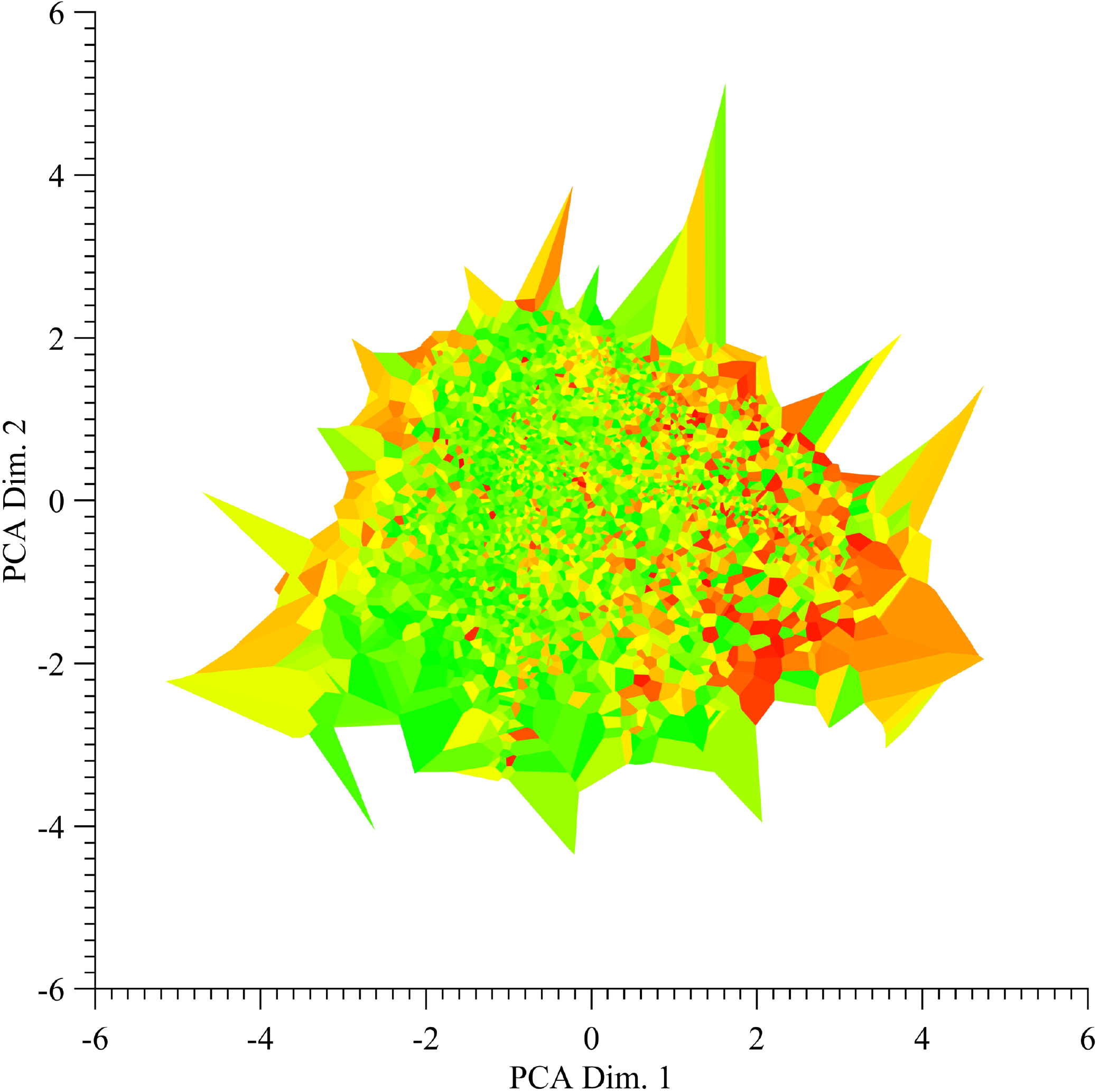}
	\label{fig:rmseLIN}%
	}
	\hfil
	\subfloat[Power forecasting model 3.]{
	\includegraphics[width=0.235\textwidth]{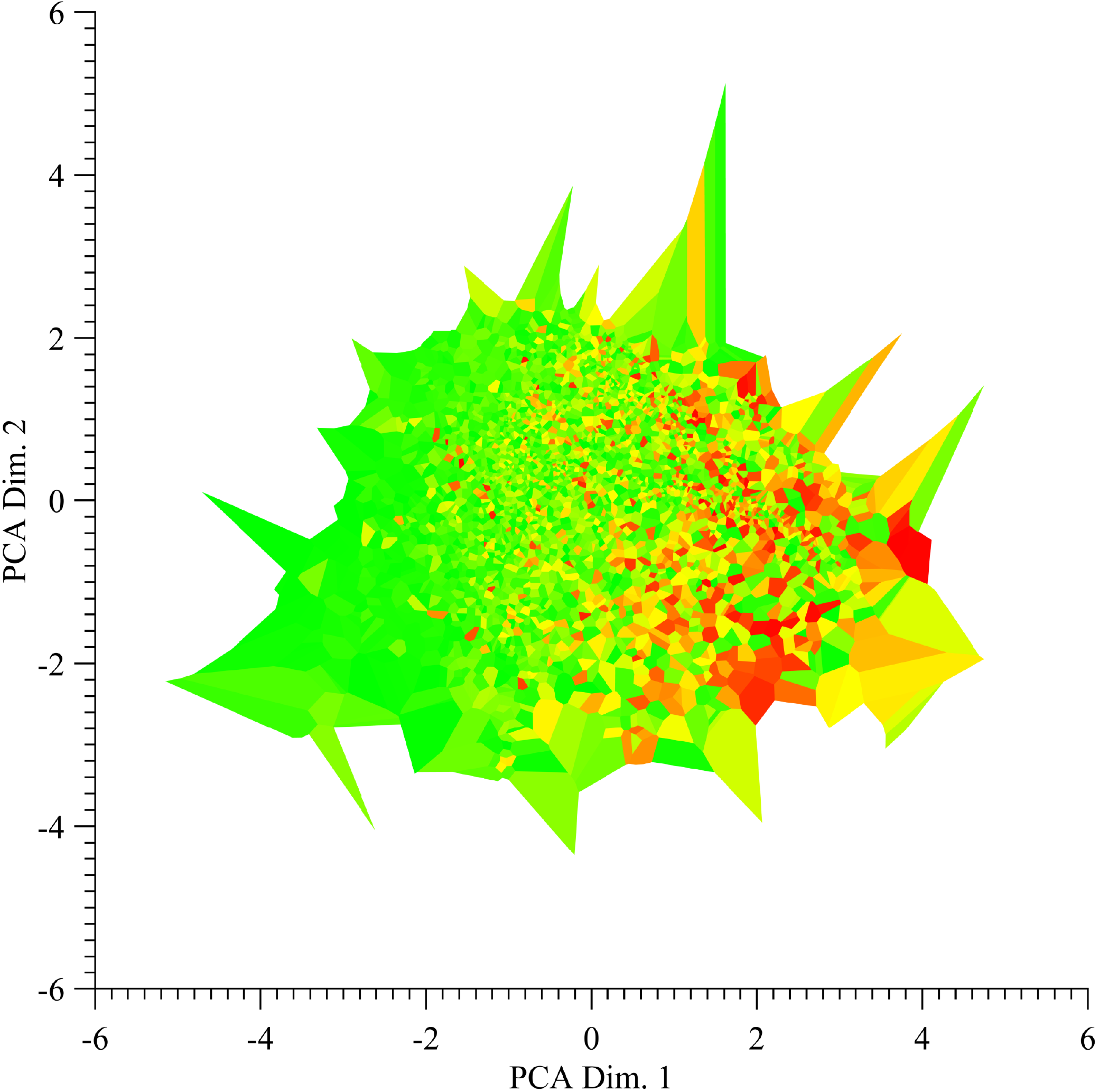}
	\label{fig:rmseKNN}%
	}
    \hfil
	\subfloat[Ensemble model.]{
	\includegraphics[width=0.235\textwidth]{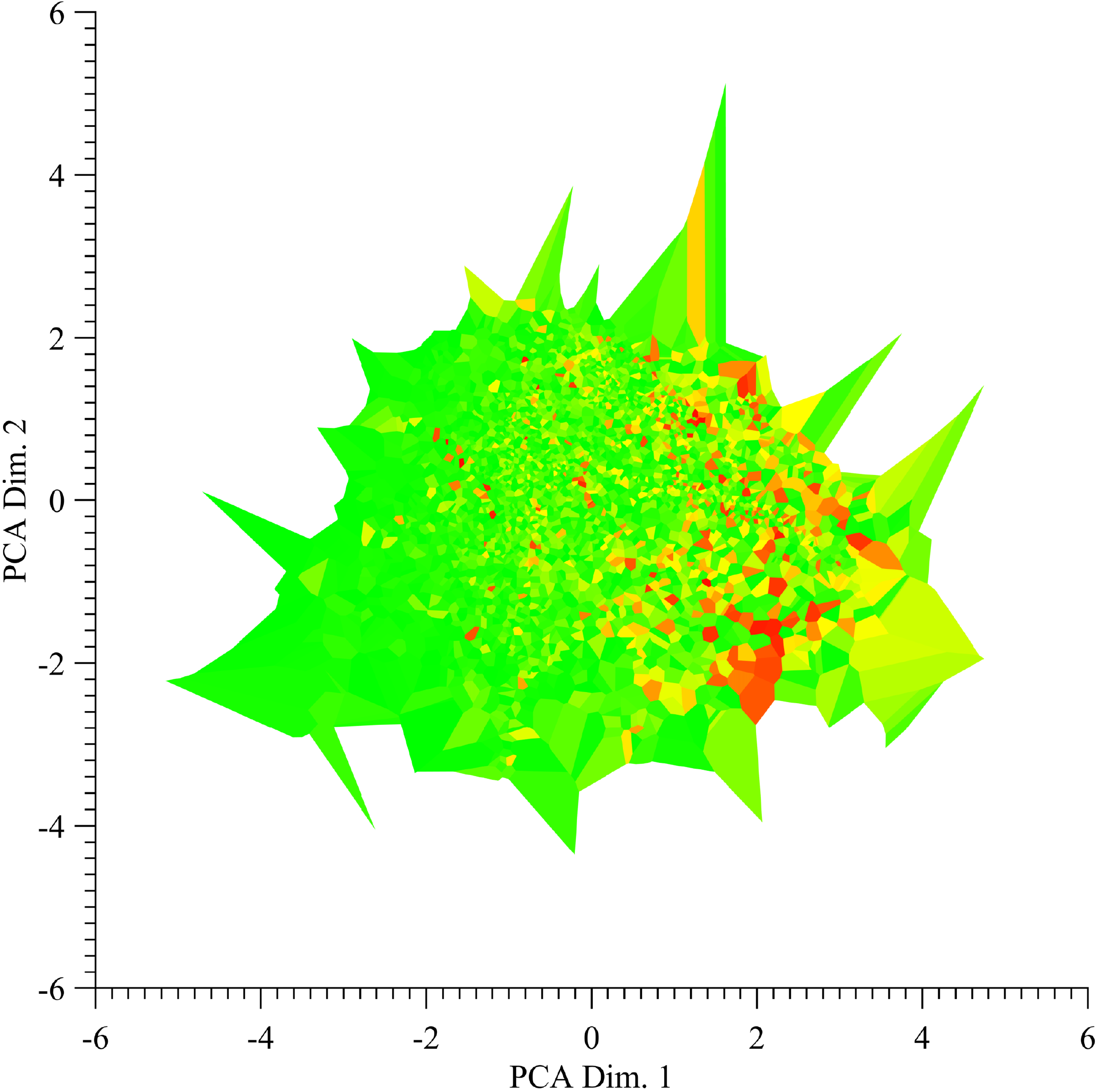}
	\label{fig:rmseENS}%
	}
	\caption{Figs. \ref{fig:rmseANN}--\ref{fig:rmseKNN} show the local error distribution of a number of power forecasting models.
	The NWPs are projected onto the two most important principal components to better visualize the high-dimensional forecast. The local quality is indicated by the colors, where green means low error, red indicates high error.
    Using the proposed local soft gating technique (Section~\ref{sec:loc}), a local soft gating for each algorithm is created. 
    Using this form of weighting, the local quality can be increased, as shown by example of the training set in Fig.~\ref{fig:rmseENS}.
    }
	\label{fig:blurriness}%
\end{figure*}

\subsection{Global Weighting}
\label{sec:glob}

The global weights $w^{(\psi)}_g, w^{(\varphi|\psi)}_g$ are fixed weights which are determined during ensemble training.
The proposed weighting technique aims at weighting the ensemble members according to their performance using {coopetitive soft gating}. A simple yet popular measure to assess the quality of an ensemble member can, for instance, be the root mean squared error (RMSE) computed by
\begin{eqnarray}
\label{eq:rmse}
  R^{(\varphi|\psi)} &=& \sqrt{\frac{1}{N} \sum_{n=1}^N (e^{(\varphi|\psi)}_n)^2}, \\
  \label{eq:rmse2}
  e^{(\varphi|\psi)}_n &=& \hat{y}^{(\varphi|\psi)}_n - o_n
\end{eqnarray}
on a data set containing $N$ samples. The $o_n$ are actual power measurements corresponding to a forecast $\hat{y}^{(\varphi|\psi)}_n$ computed on weather forecasting model $\psi$ and power forecasting model $\varphi$.

The global weight $w^{(\varphi|\psi)}_{g}$ can be computed from the coopetitive soft gating formula
\begin{eqnarray}
  w^{(\varphi|\psi)}_{g} &=& \varsigma_\eta(\bm{g}^{(\psi)}, R^{(\varphi|\psi)}), \\
  \bm{g}^{(\psi)} &=& \big( R^{(1|\psi)},\denseDots,R^{(\varphi|\psi)},\denseDots, R^{(\Phi|\psi)} \big),
\end{eqnarray}
where $\varsigma_\eta$ is the coopetitive soft gating function of Eq.~(\ref{eq:weightingFormulaGeneral}) for a power forecasting model $\varphi$ computed on weather forecasting model $\psi$.

The global forecasting ability of a weather forecasting model $\psi$ can be observed only indirectly as actual weather measurements typically are not available.
As an estimate, the overall quality of a weather forecasting model can be determined using
the average quality of all power forecasting models for the particular weather forecasting model, i.e.,
\begin{eqnarray}
\label{eq:wmglobWeighting}
  w^{(\psi)}_g &=& \varsigma_\eta(\bm{p},p^{(\psi)}), \\
  p^{(\psi)} &=& \frac{1}{\Phi} \sum_{\varphi=1}^{\Phi} {R}^{(\varphi|\psi)}, \\
  \bm{p} &=& \big( p^{(1)}, \denseDots, p^{(\psi)} ,\denseDots, p^{(\Psi)} \big).
\end{eqnarray}

An example for the influence of the global weighting term for a number of power forecasting models is given in Section~\ref{sec:coopExample}.
This weighting term is computed once during {ensemble training} and can then be reused for every forecast.

\subsection{Local Soft Gating}
\label{sec:loc}

The second weighting term depends on the values of the current NWP, i.e., the weighting is realized depending on the particular characteristics of the weather situation.
The NWP forecast can be seen as a point in a feature space which characterizes the weather situation.
The basic assumption is that both, weather and power forecasting models, may have strengths and weaknesses in varying areas of the feature space. 
This is due to the fact that different power forecasting algorithms yield different errors in certain areas of the feature space due to structure, data, or parameter diversity effects.
In particular for sparsely covered areas of the feature space (e.g., storms for wind turbines, or Sahara dust for photovoltaic plants), this effect
may become more prominent.
Different NWP forecasts of certain weather forecasting models may also have a different precision depending on the particular situation.
Using coopetitive soft gating, we aim to exploit the advantages of each model in particular observed situations during model training.

In order to obtain local weights $w^{(\psi)}_l$ and $w^{(\varphi|\psi)}_l$, the neighborhood of a weather forecast $\mathbf{x}_{t_k|t}$ has to be assessed.
Similar historic weather situations are found with respect to a (historic) data set tuple $\mathbf{X}_{\text{H}} = \big(\mathbf{x}_1, \ldots, \mathbf{x}_N \big)$ with $\mathbf{x}_n \in \mathbb{R}^{D}$,
which is used during ensemble training.
The proposed ensemble algorithm is able to work with an arbitrary local quality assessment technique.
Here we demonstrate the application with a simple nearest neighbor technique.
Other techniques for assessing locality, such as multi-linear interpolation, are investigated, e.g., in \citep{Gensler2016ensemble}.

%
%

A simple yet effective technique for locality assessment is a nearest neighbor algorithm.
In order to assess the local quality of a forecast $\mathbf{x}^{(\psi)}_{t_k|t}$, its $C$ nearest neighbors
are searched in $\mathbf{X}_{\text{H}}$ in the way
\begin{equation}
  \bm{\alpha}^{(\psi)} = \text{knn}(\mathbf{x}^{(\psi)}_{t_k|t},\mathbf{X}_{\text{H}},C)~,~~~\bm{\alpha}^{(\psi)} \in \mathbb{N}^C,
\end{equation}
where $\bm{\alpha}^{(\psi)}$ is a set containing the indices of the $C$ nearest neighbors.
Here, we use the Euclidean distance as distance metric on standardized input dimensions, though the use of more advanced distance metrics may further improve the local quality assessment.
The average local quality can be assessed using
\begin{equation}
\label{eq:knnQuality}
  q^{(\varphi|\psi)} = \frac{1}{c} \sum_{a \in \bm{\alpha}^{(\psi)}} |e_a^{(\varphi|\psi)}|,
\end{equation}
where $e_a^{(\varphi|\psi)}$ is the error computed using Eq.~(\ref{eq:rmse2}) of the item at index $a$ in the historic set using the forecast of ensemble member with power forecasting model $\varphi$ computed on weather forecasting model $\psi$.
From this local error score $q^{(\varphi|\psi)}$, the weight $w^{(\varphi|\psi)}_l$ is computed for each power forecasting model using coopetitive soft gating of Eq.~(\ref{eq:weightingFormulaGeneral}) in the form 
\begin{eqnarray}
\label{eq:localityFMdl}
  w^{(\varphi|\psi)}_l &=&  \varsigma_\eta(\mathbf{q}^{(\psi)},q^{(\varphi|\psi)}), ~~~\\
  \mathbf{q}^{(\psi)} &=& \big( q^{(1|\psi)},\denseDots, q^{(\varphi|\psi)}, \denseDots, q^{(\Phi|\psi)} \big),
\end{eqnarray}
where each value of $q^{(\varphi|\psi)}$ is computed using Eq.~(\ref{eq:knnQuality}). 
In the same fashion as for the global weighting (Eq.~(\ref{eq:wmglobWeighting})),
the relative quality for each weather forecasting model $\psi$ is estimated indirectly using all available power forecasting models, i.e.,
\begin{eqnarray}
  w^{(\psi)}_l &=&  \varsigma_\eta(\bar{\mathbf{q}},\bar{q}^{(\psi)}), ~\text{where}\\
  \bar{q}^{(\psi)} &=& \frac{1}{\Phi} \sum_{\varphi=1}^{\Phi}  q^{(\varphi|\psi)}, \\
  \bar{\mathbf{q}} &=& \big( \bar{q}^{(1)},\denseDots, \bar{q}^{(\psi)},\denseDots,\bar{q}^{(\Psi)}  \big).
\end{eqnarray}

Figs.~\ref{fig:rmseANN} -- \ref{fig:rmseKNN} show the local quality of a number of power forecasting models in Voronoi diagrams, where green
represents areas of low error and red color indicates areas with high error.
The axes are given by the two most important principal components in order to better
visualize the $D$-dimensional NWP feature space.
Using the locality assessment technique of  Eq.~(\ref{eq:localityFMdl}), each model is weighted depending on the position in the feature space in a
way that reduces the overall error in the ensemble.
Fig.~\ref{fig:rmseENS} shows an example of the resulting ensemble error.
It should be kept in mind that in the shown case, the improvement for the training data set is displayed.
An example for the development of the local weights in a forecasting time period is shown in Section~\ref{sec:coopExample}.
This weighting term has to be computed during ensemble application for every NWP.

An advantage of the knn technique is that no model training is required (in the basic form if no feature subspace is selected or feature weighting is applied). However, as the data set $\mathbf{X}_\text{H}$ serves as basis for the locality assessment, it has to be searched in every iteration, which usually does not scale optimally if no search heuristics are being employed.
The knn approach is therefore particularly useful for smaller data sets.
In \citep{Gensler2016ensemble}, a technique for locality assessment based on multi-linear interpolation is introduced which does require a training phase. However, during model application it no longer requires the data set $\mathbf{X}_\text{H}$.
This technique is therefore well-suited for larger data sets.
Regarding ensemble forecasting quality, both approaches turned out to behave similar.

\subsection{Lead Time-Dependent Soft Gating}
\label{sec:dt}

The lead time-dependent weighting components $w^{(\psi)}_{k}, w^{(\varphi|\psi)}_{k}$ factor in the quality development of a model for each lead time $k$.
The idea is to weigh models according to their lead time-dependent performance.
In the area of power forecasting, a prominent example for approaches with time step dependent performance is the persistence method, which performs well on very short lead times only.

The idea is to create a weight per lead time $k$ by evaluating the quality differences of a number of coopetitive models.
For the creation of this form of weighting
a training data set for a particular lead time $k$ -- for which a number of forecasts $\hat{\mathbf{y}}^{( k,\varphi|\psi)} \in \mathbb{R}^{N'}$ are created using weather forecasting model~$\psi$ and power forecasting model $\varphi$~--
can be denoted as
\begin{equation}
  \hat{\mathbf{y}}^{( k,\varphi|\psi)} = \big( \hat{y}_{t_1+ \Delta k| t_1}^{(\varphi|\psi)},\hat{y}_{t_{2}+ \Delta k| t_{2}}^{(\varphi|\psi)},\denseDots,\hat{y}_{t_{N'}+ \Delta k| t_{N'}}^{(\varphi|\psi)} \big),
\end{equation}
where $N'$ is the number of evaluated elements with the currently evaluated lead time $k$.
The estimated quality for a particular lead time can then be created using an error metric, e.g., based on the RMSE
\begin{equation}
  {R}^{(\varphi|\psi)}_{ k} = \sqrt{\frac{1}{N'} \sum_{n=1}^{N'} ({\hat{\mathbf{y}}}^{( k,\varphi|\psi)}_{n} - o_{n})^2}.
\end{equation}
The quality of the particular forecasting time step $k$ in relation to other forecasting time steps of the same model can then denoted as
\begin{equation}
  r^{(\varphi|\psi)}_{ k} = \frac{{R}^{(\varphi|\psi)}_{ k}}{\frac{1}{k_{\text{max}}-k_{\text{min}}+1} \sum_{k^*=k_{\text{min}}}^{k_{\text{max}}} {R}^{(\varphi|\psi)}_{k^*}}.\\
\end{equation}
%
%
Then, the weighting factor $w^{(\varphi|\psi)}_{ k}$ is computed for each forecasting time step using the generalized coopetitive soft gating formula of Eq.~(\ref{eq:weightingFormulaGeneral}) in relation to other members of the ensemble
\begin{eqnarray}
\label{eq:dtFMdl}
  w^{(\varphi|\psi)}_{k} &=&  \varsigma_\eta(\mathbf{r}^{(\psi)}_{ k},r^{(\varphi|\psi)}_{ k}), ~\text{where}\\
  \mathbf{r}^{(\psi)}_{k} &=& \big( r^{(1|\psi)}_{ k},\denseDots, r^{(\varphi|\psi)}_{ k},\denseDots,r^{(\Phi|\psi)}_{ k} \big).
\end{eqnarray}
Again, the time-dependent weather forecasting model qualities are estimated using the overall power forecasting models with
\begin{eqnarray}
  w^{(\psi)}_{ k} &=&  \varsigma_\eta(\bar{\mathbf{r}}_{ k},\bar{r}^{(\psi)}_{ k}), ~\text{where}\\
  \bar{r}^{(\psi)}_{ k} &=& \frac{1}{\Phi} \sum_{\varphi=1}^{\Phi}  r^{(\varphi|\psi)}_{ k}, \\
  \bar{\mathbf{r}}_{ k} &=& \big(\bar{r}^{(1)}_{ k},\denseDots, \bar{r}^{(\psi)}_{ k} ,\denseDots,\bar{r}^{(\Psi)}_{ k} \big).
\end{eqnarray}
%
%
In case there is little data available for the training process, a smoothing over weights in neighboring lead times can be applied in order to avoid
noisy weights.
An example of the effect of this form of weighting is described in Section~\ref{sec:coopExample}.
This weighting term is computed once during {ensemble training} for every lead time and can then be reused for every forecast.

\begin{figure*}[ptb]
	\centering
	\subfloat[Development of global weighting $w^{(\varphi|\psi)}_g$ over time. ANN performs best, followed by the linear regression. The global weighting of the persistence method is small.]{%
	\includegraphics[width=0.44\textwidth]{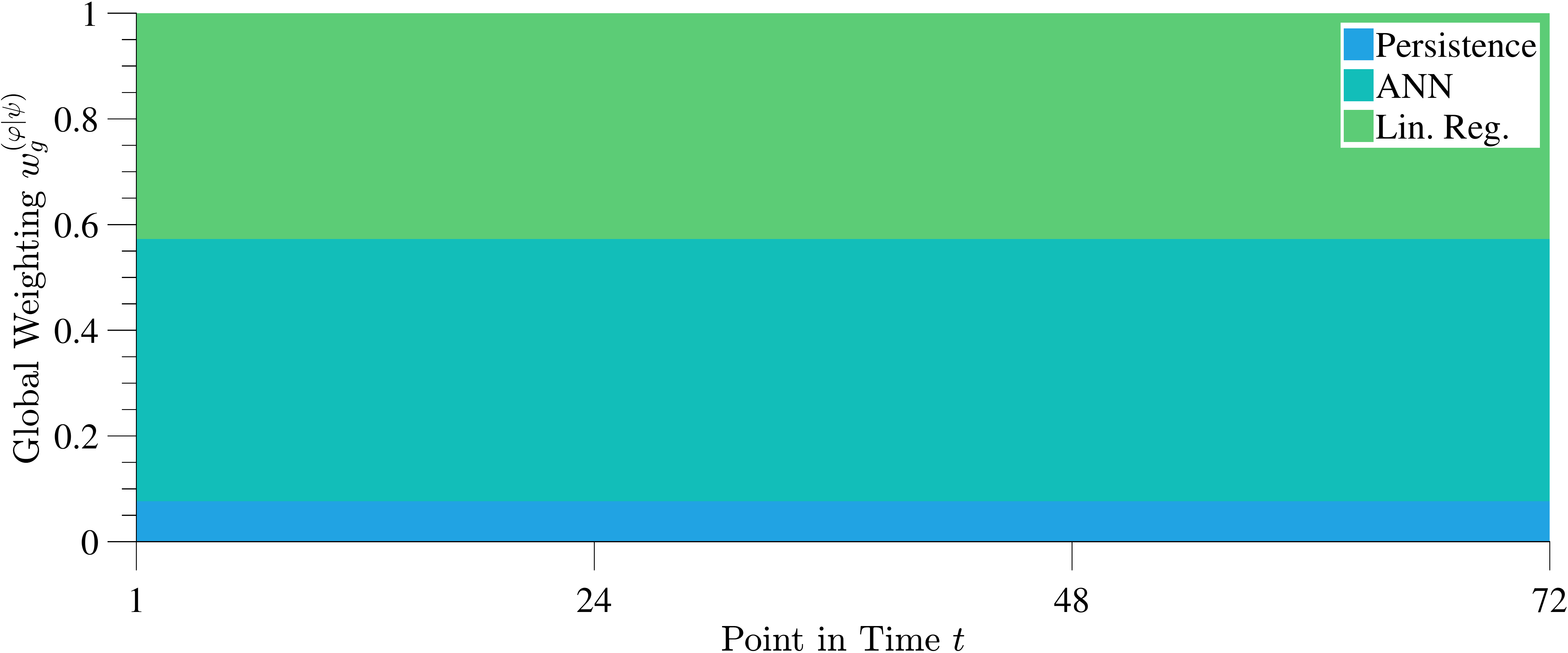}
	\label{fig:smGlob}%
	}
	\hfil
\subfloat[Development of local weighting $w^{(\varphi|\psi)}_l$ over time. As the particular weather situation is a fairly standard situation (and thus, it is relatively well represented in the
    data set), there is little observable local weighting in this particular case.]{
	\includegraphics[width=0.44\textwidth]{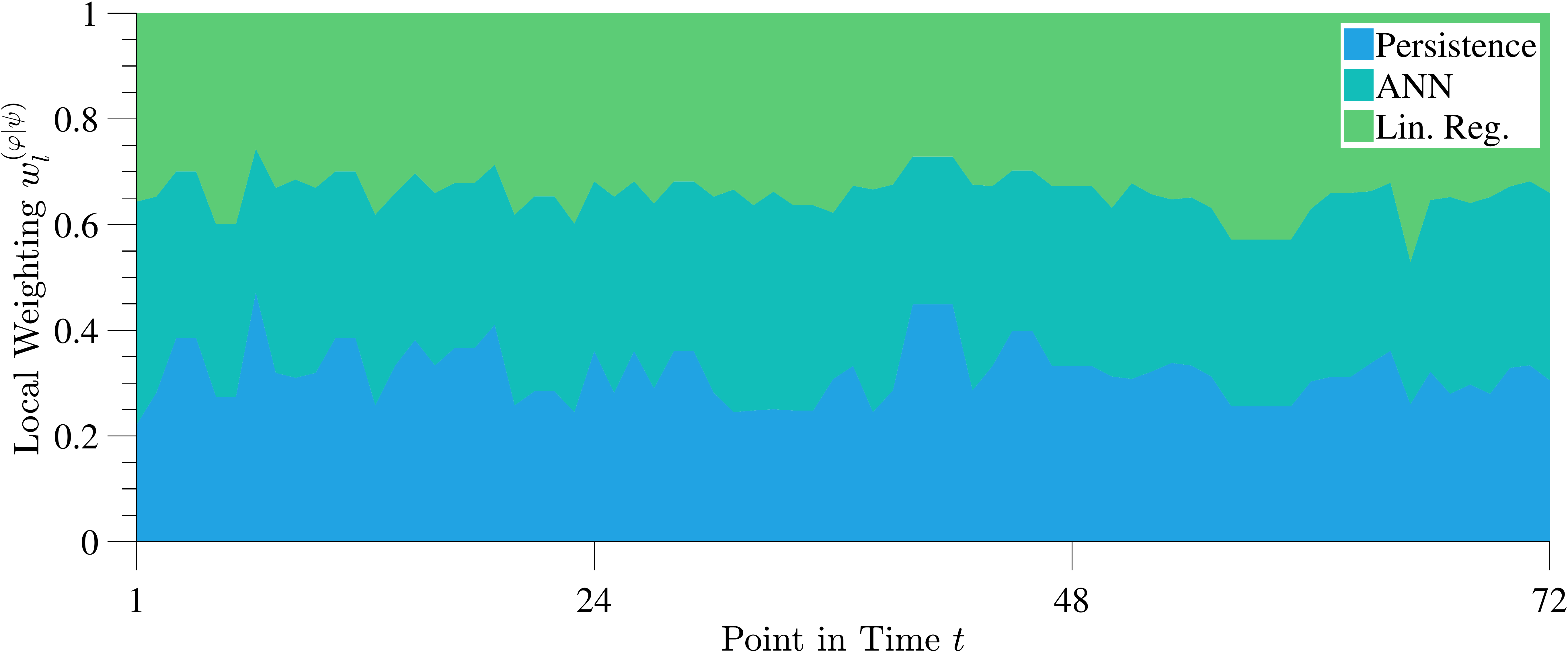}
	\label{fig:smLoc}%
	}
	\hfil
	\subfloat[Development of lead time-dependent weighting $w^{(\varphi|\psi)}_{ k}$ over time. Note the different time-scale ($24$ h). Though the global performance $w^{(\varphi|\psi)}_g$ of the persistence method is low, it performs
    strongly on short time horizons. 
    ]{%
	\includegraphics[width=0.44\textwidth]{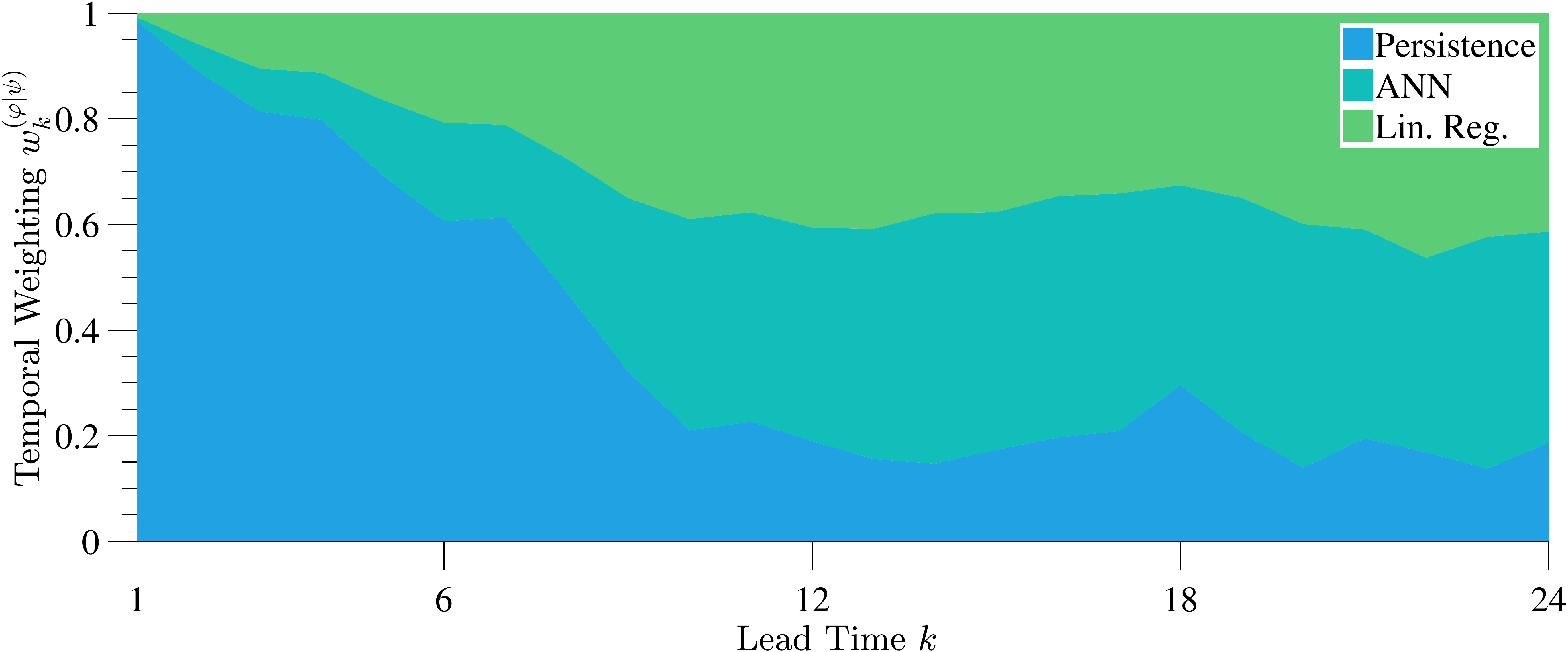}
	\label{fig:smDt}%
	}
	\hfil
	\subfloat[Combined $w^{(\varphi|\psi)}$, consisting of $w^{(\varphi|\psi)}_g$, $w^{(\varphi|\psi)}_{ k}$, and $w^{(\varphi|\psi)}_l$, over time. As can be seen in this example,
    though the overall quality of the persistence method is low, it has high impact in short time horizons.]{
	\includegraphics[width=0.44\textwidth]{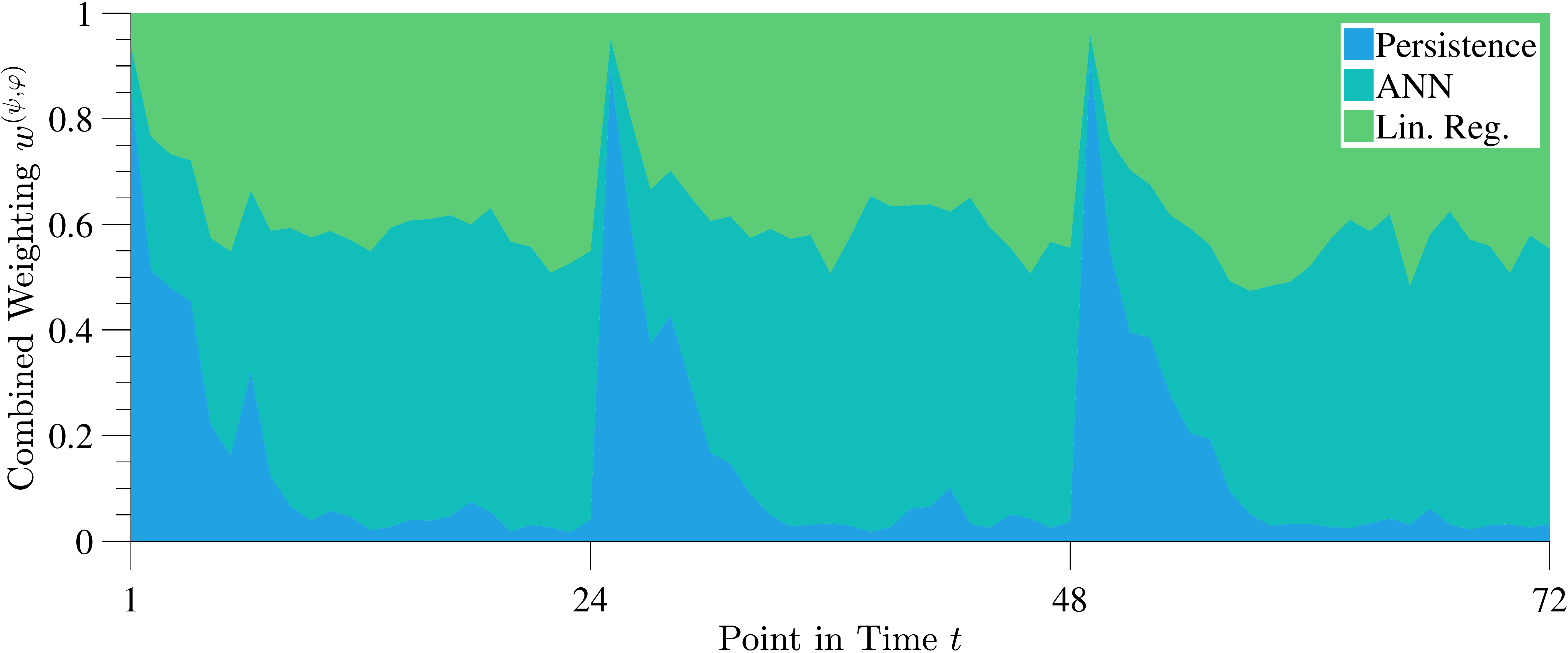}
	\label{fig:smAll}%
	}
	\hfil
	\subfloat[Forecasting example using combined weighting of the CSGE technique. On delivery of a new NWP each $24$ h, the weighting of the persistence method is high in this example.]{
	\includegraphics[width=0.6\textwidth]{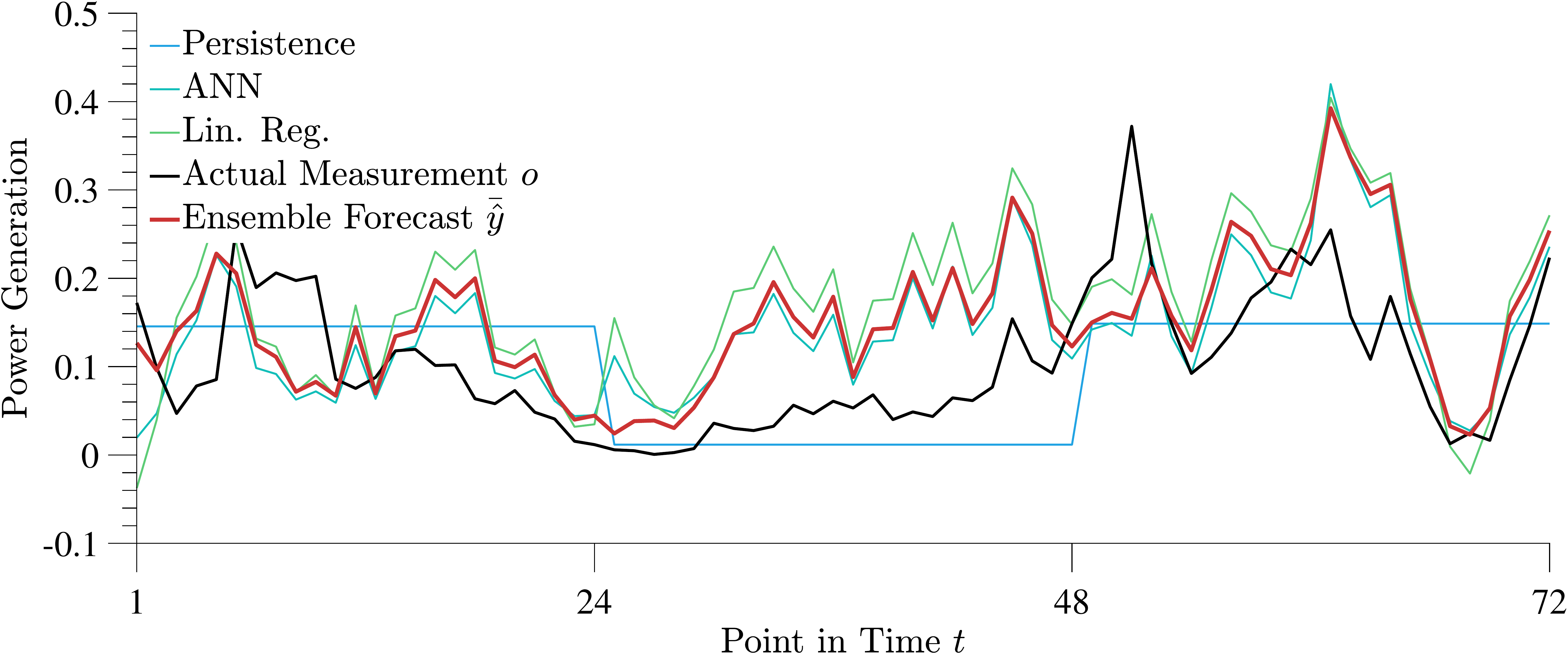}
	\label{fig:smFc}%
	}
	\caption{Example of the weight combination of the proposed CSGE ensemble technique. In the example, an intraday forecast is performed using a single weather forecasting model and three power forecasting models. New weather data are incorporated every $24$ h.
    For the intraday forecast, a persistence method is combined with an artificial neural network and a linear regression model. During ensemble training, the coopetitive soft-gating parameters are optimized so that the depicted weighting emerges for an example weather situation over three days.
    The overall weight development $w^{(\psi,\varphi)}$ (Fig.~\ref{fig:smAll}) is composed of the three weighting terms $w^{(\varphi|\psi)}_g$ (Fig.~\ref{fig:smGlob}), $w^{(\varphi|\psi)}_l$ (Fig.~\ref{fig:smLoc}), and $w^{(\varphi|\psi)}_{ k}$ (Fig \ref{fig:smDt}). The overall forecast is shown in Fig.~\ref{fig:smFc}.}
	\label{fig:intradayfig}%
\end{figure*}

\begin{figure*}[htb]
	\centering
	\subfloat[Multi-model CSGE consisting of $3$ weather forecasting models, each predicted with $2$ power forecasting models. ]{%
	\includegraphics[width=0.49\textwidth]{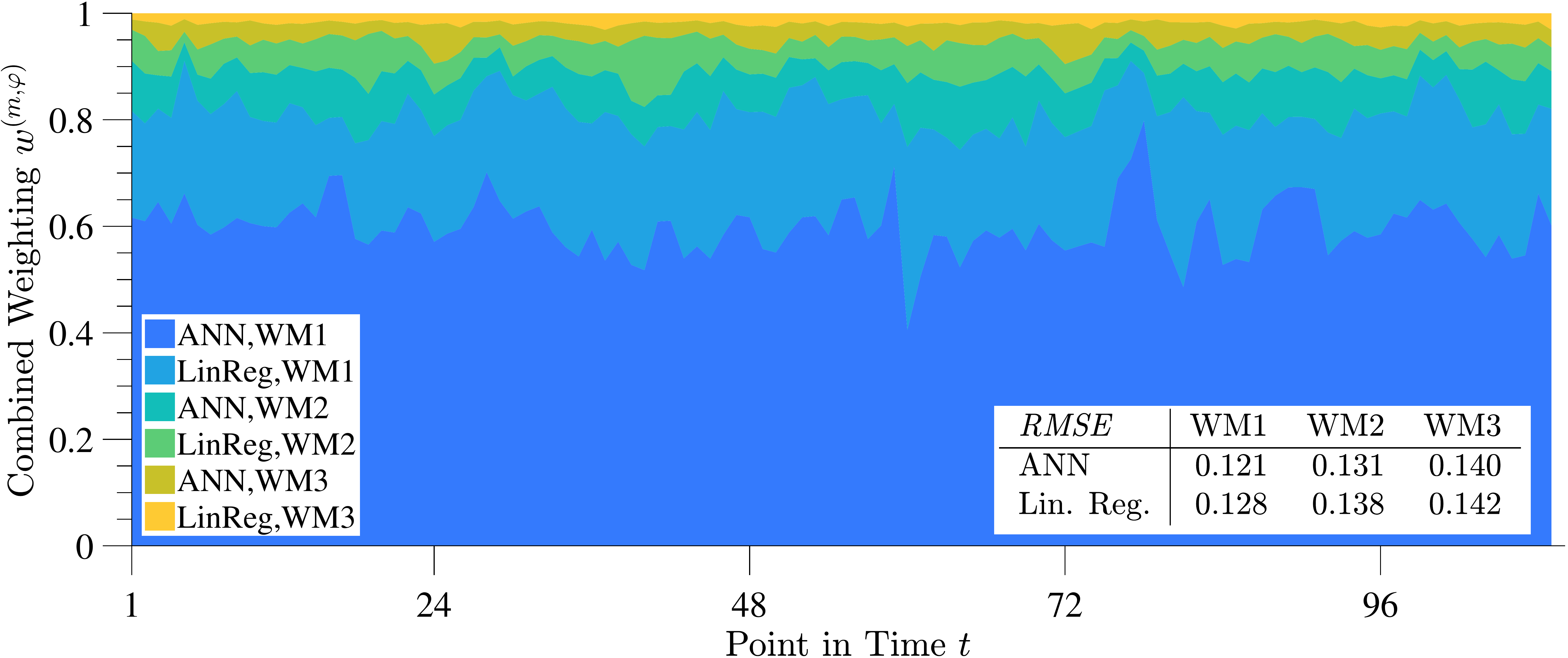}
	\label{fig:mmAll}%
	}
	\hfil
	\subfloat[Multi-model CSGE forecast with a total of $J=6$ ensemble members.]{%
	\includegraphics[width=0.49\textwidth]{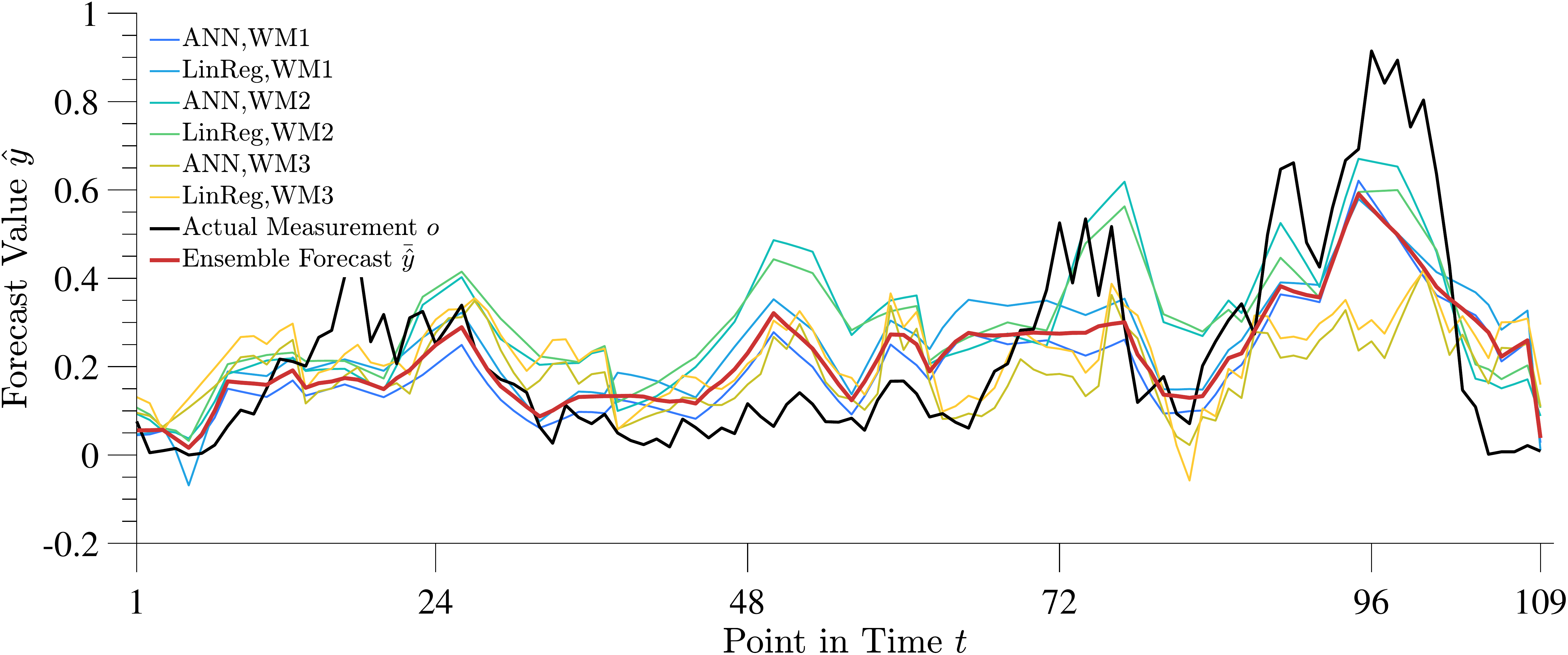}
	\label{fig:mmFc}%
	}
	\caption{Example of weighting combination for a day-ahead forecast using three weather forecasting models (multi-model ensemble) and the proposed CSGE technique. Each model has different forecasting quality (where it can be observed that WM1 has the highest quality and WM3 performs worst).
    The overall weighting for each weather / power forecasting model combination is shown in Fig.~\ref{fig:mmAll}.
    The result is in line with the expectation when performing the prediction using a single weather and power forecasting model combination, as can be seen from the table in the figure. Values denote RMSE on normalized power values.
    The overall forecast is shown in Fig.~\ref{fig:mmFc}.
    Depending on the particular situation, the ensemble adapts the model weights.}
	\label{fig:mmWeightingExample}%
\end{figure*}

\subsection{Model Fusion and Ensemble Training}
\label{sec:overallmdl}

As stated previously, the overall weighting $w^{(j)}$ of each ensemble member $j$ is computed using Eq.~(\ref{eq:overallweighting}).
The main parameter of the \emph{coopetitive soft gating ensemble} (CSGE) algorithm is the hyperparameter $\eta$ of the coopetitive soft gating formula.
Depending on the forecasting task and the data set, the appropriate value of $\eta$ may differ.
Furthermore, the value of $\eta$ for each weighting aspect (global weighting, local soft gating, time-dependent soft gating)
may vary. The value of $\eta$ therefore should be chosen independently for each weighting aspect.

%
%

In principle, the total number of $\eta$-parameters $S$ are the three aspects (global, local, temporal) for each weather forecasting model. Additionally, for each of the $\Psi$ weather forecasting models, the three weighting aspects for the corresponding power forecasting models have to be computed, which adds up to
\begin{equation}
S = \underbrace{3 \cdot \Psi}_{{\substack{\text{power} \\ \text{forecasting} \\ \text{models}}}} + \underbrace{3}_{\substack{\text{weather} \\ \text{forecasting} \\ \text{models}}}.
\end{equation}
Assuming the same number and types of power forecasting models for each weather forecasting model, the weighting parameter can be treated identically for each weather forecasting model. Therefore, the number of optimization parameters can be
reduced to
\begin{equation}
S = \underbrace{3}_{{\substack{\text{power} \\ \text{forecasting} \\ \text{models}}}} + \underbrace{3}_{\substack{\text{weather} \\ \text{forecasting} \\ \text{models}}} = 6
\end{equation}
parameters.
The tuple of coopetitive soft gating parameters $\bm{\eta} = (\eta_1, \ldots, \eta_s, \ldots, \eta_S )$ can then be optimized using an arbitrary optimization algorithm solving the problem
\begin{equation}
\label{eq:optimization}
\begin{aligned}
& \underset{\bm{\eta}}{\text{minimize}} & & \frac{1}{N} \sum_{n=1}^N [o_n -  f_{\text{CSGE}}(\mathbf{x}_{n},\bm{\eta}) ]^2 + \zeta \cdot \sum_{s=1}^{S} {\eta}_s, \\
& \text{where} & &  f_{\text{CSGE}}(\mathbf{x}_{n},\bm{\eta}) = \sum_{j=1}^{J} w^{(j)}_{\bm{\eta}} \cdot \hat{y}^{(j)}_{n}, \\
& \text{subject to} & & ~\text{each}~ {\eta}_s \geq 0,
\end{aligned}
\end{equation}
where $f_{\text{CSGE}}$ is the forecast of the overall CSGE forecasting function, $N$ are the overall evaluated points of a validation data set, $w^{(j)}_{\bm{\eta}}$ are the weights of a particular forecast computed using
Eq.~(\ref{eq:overallweighting}) using the hyperparameters $\bm{\eta}$, and $\zeta \geq 0$ is a regularization parameter.
In this case, a squared error is chosen for optimization, however, depending on the application, also other forms of error functions can be used.

An advantage of the proposed technique is that it is a post-processing technique. Therefore, while the single forecasts of each ensemble member $j$ are weighted differently using $w^{(j)}_{\bm{\eta}}$,
the forecasts $\hat{y}^{(j)}_{n}$ of each of the $j$ ensemble members remains constant, no matter what the value of $\bm{\eta}$ may be.
The values of $\hat{y}^{(j)}_{n}$ therefore only have to be computed once for the evaluated data set during ensemble training.
Furthermore, the single weights change gradually when varying the parameters in $\bm{\eta}$.
Consequently, we end up with a smooth (continuously differentiable) optimization function. 

\subsection{Application Examples of the CSGE Technique}
\label{sec:coopExample}

This section describes two application examples of the final CSGE algorithm.
The first example shows the application of the CSGE algorithm for intraday forecasting ($k_{\mathrm{min}} = 1$, $k_{\mathrm{max}}=24$, $\Delta = 1 h$) using a single weather forecasting model and an ensemble of three forecasting algorithms, namely an ANN, a linear regression, and a persistence forecast.
Fig.~\ref{fig:intradayfig} shows the single weighting aspects and the overall weights over time. As there exists just a single weather forecasting model, the number of weighting aspects is reduced to three.
The global weights are shown in Fig.~\ref{fig:smGlob}. The algorithm weighs the single algorithms according to their expected quality (ANN best, persistence worst). These weights remain constant over time.
The local weights are detailed in Fig.~\ref{fig:smLoc}. In this particular case, all local weights are similar.
The lead time-dependent weights are shown in Fig.~\ref{fig:smDt}. Note the different horizontal axis, which is the lead time in this case. As is to be expected, the persistence method works well on very short time horizons, but
quickly loses quality in comparison to the other two approaches. 
The combination of the three weighting aspects is visualized in Fig.~\ref{fig:smAll}. As can be seen, on delivery of new NWP forecasts every $24 h$, the influence of the lead time-dependent persistence technique is high.
An overall forecast is shown in Fig.~\ref{fig:smFc}.

The second example shows a multi-model forecast using three weather forecasting models for a day-ahead forecast, each of which is predicted using two power forecasting models.
This example is visualized in Fig.~\ref{fig:mmWeightingExample}.
The overall weighting over time is shown in Fig.~\ref{fig:mmAll}.
Regarding the weather forecasting models, the first weather forecasting model ``WM1'' (which is the ECMWF IFS model) has the highest influence, while the other two weather forecasting models have lower weights.
This, again, meets the expectation, as can be seen from the table in Fig.~\ref{fig:mmAll}, which shows the overall RMSE when performing the forecast on a single weather / power forecasting model combination.
The model with the lowest RMSE error gets the highest weight.
Also, the quality difference between the power forecasting models is reflected in the weighting.
As there are no weather or power forecasting models which are designed for a different forecasting time period (unlike in the first example), the overall weighting differences over time are not as drastic as in the first example.
The forecast which is created using the weights determined by the CSGE is shown in Fig.~\ref{fig:mmFc}.

\section{Experimental Results}
\label{sec:result}

This section investigates the performance of the proposed CSGE technique in comparison to a number of
state of the art approaches. We evaluate the algorithms on $45$ data sets which are described in Section~\ref{sec:dataset}.
The experimental setup is described in Section~\ref{sec:experimentSetup}.
We examine the proposed power forecasting model using a single weather forecasting model (Section~\ref{sec:smResults}), and using
multiple weather forecasting models for day-ahead forecasting (Section~\ref{sec:mmResults}), as well as for intraday forecasts (Section~\ref{sec:idResults}). Finally, a detailed comparison of the performance gains when using multiple
weather forecasting models for both, day-ahead and intraday forecasts, is detailed in Section~\ref{sec:mmGains}.
A discussion of applicability is performed in Section~\ref{sec:discussion}.

\begin{figure}[tb]
\centering
\includegraphics[width=.47\textwidth]{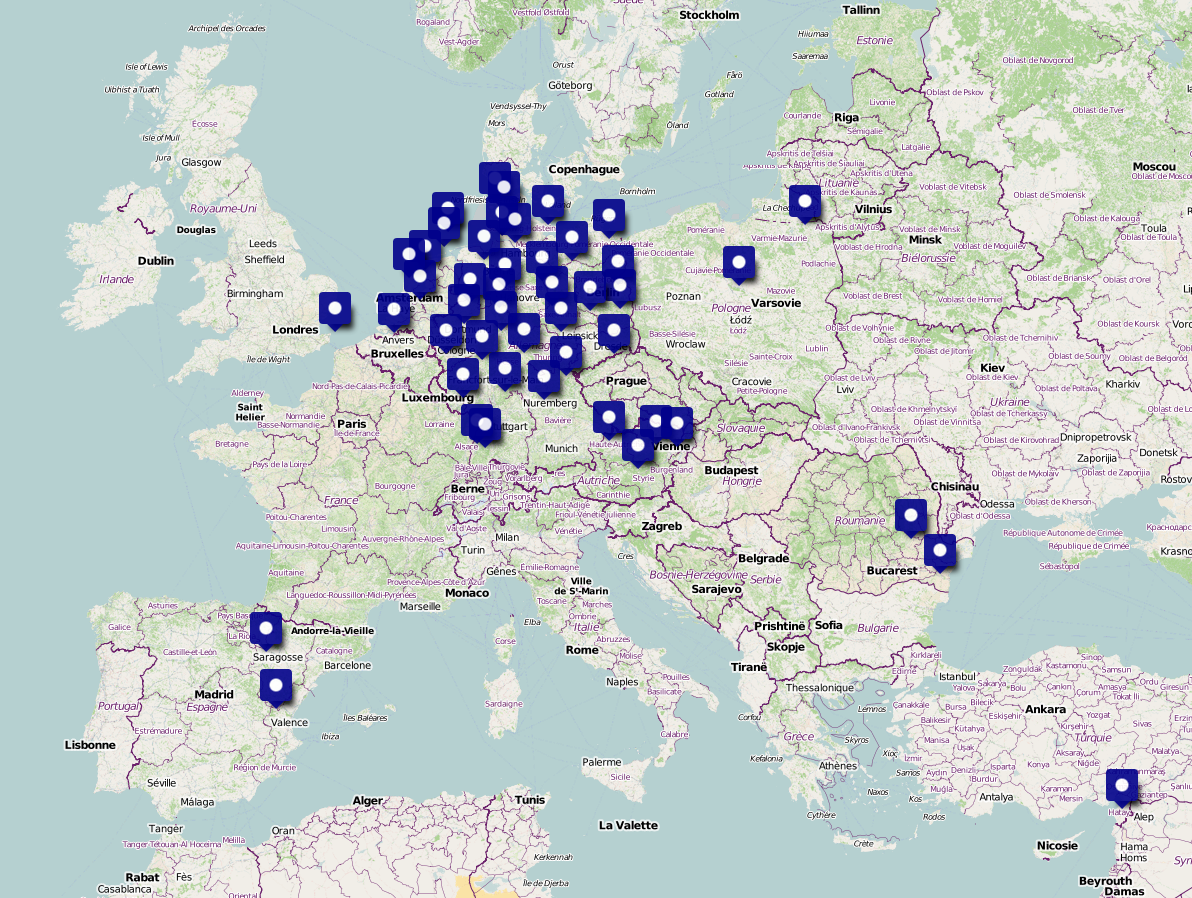}
\vspace{-6pt}
\caption{Locations of the wind farms of the \emph{EuropeWindFarm} collection, see Section~\ref{sec:dataset} for a data set description.}
\label{fig:europe}
\end{figure}

\subsection{Data Set Description}
\label{sec:dataset}

The data sets used for the evaluation are the publicly available data sets of the \emph{EuropeWindFarm} collection \citep{Gensler2016} using multiple weather forecasting models.
The $45$ data sets contain the weather forecasts and power measurements of two consecutive years of both onshore and offshore wind farms.
The data sets contain the following items:
\begin{itemize}[noitemsep]
	\item \textit{Time Stamp} of the forecast / power measurement,
	\item \textit{Lead time} from the forecasting origin,
	\item \textit{Wind Speed} in $100~m$ height,
	\item \textit{Wind Speed} in $10~m$ height,
	\item \textit{Wind Direction (zonal)} in $100~m$ height,
	\item \textit{Wind Direction (meridional)} in $100~m$ height,
	\item \textit{Air Pressure} forecast,
	\item \textit{Air Temperature} forecast, and
	\item \textit{Power Generation} of the wind farm.
\end{itemize}
The power generation time series are normalized with respect to the nominal capacity of each power plant to enable a scale-free comparison.
All weather input parameters are normalized to in the interval $[0,1]$.
The data has been filtered to eliminate erroneous measurements.

\begin{table*}[htb]
  \centering
  \scriptsize
  \caption{Performance comparison regarding RMSE and the $R^2$ score of a number of power forecasting algorithms and ensembles on the EuropeWindFarm data sets, using a single weather forecasting model for day-ahead forecasting. The color coding indicates the quality of each wind farm and power forecasting algorithm from high quality (green) to low quality (red). Bold text highlights the best achieved score for each evaluated data set. }
  \resizebox{\textwidth}{!}{\begin{minipage}{1.2\textwidth}
  \centering
\begin{tabular}{r|rrrrrrr|rrrrrrr}
    \cellcolor[rgb]{1,1,1}      & \multicolumn{7}{c|}{RMSE}                             & \multicolumn{7}{c}{$R^2$} \\
    \cellcolor[rgb]{1,1,1}      & \cellcolor[rgb]{1,1,1}ANN   & \cellcolor[rgb]{1,1,1}Lin. Reg. & \cellcolor[rgb]{1,1,1}LSBoost & \cellcolor[rgb]{1,1,1}Bag.  & \cellcolor[rgb]{1,1,1}CE    & \cellcolor[rgb]{1,1,1}CSGE-K & \cellcolor[rgb]{1,1,1}CSGE-I & \cellcolor[rgb]{1,1,1}ANN   & \cellcolor[rgb]{1,1,1}Lin. Reg. & \cellcolor[rgb]{1,1,1}LSBoost & \cellcolor[rgb]{1,1,1}Bag.  & \cellcolor[rgb]{1,1,1}CE    & \cellcolor[rgb]{1,1,1}CSGE-K & \cellcolor[rgb]{1,1,1}CSGE-I \bigstrut[b]\\
    \hline
    \cellcolor[rgb]{1,1,1}wf1   & \cellcolor[rgb]{0.768627450980392,0.854901960784314,0.501960784313725}0.111 & \cellcolor[rgb]{0.819607843137255,0.870588235294118,0.505882352941176}0.118 & \cellcolor[rgb]{0.772549019607843,0.854901960784314,0.501960784313725}0.112 & \cellcolor[rgb]{0.745098039215686,0.847058823529412,0.501960784313725}0.109 & \cellcolor[rgb]{0.741176470588235,0.847058823529412,0.501960784313725}\textbf{0.108} & \cellcolor[rgb]{0.737254901960784,0.843137254901961,0.501960784313725}\textbf{0.108} & \cellcolor[rgb]{0.737254901960784,0.843137254901961,0.501960784313725}\textbf{0.108} & \cellcolor[rgb]{0.913725490196078,0.898039215686275,0.513725490196078}0.714 & \cellcolor[rgb]{0.996078431372549,0.905882352941176,0.513725490196078}0.683 & \cellcolor[rgb]{0.949019607843137,0.905882352941176,0.517647058823529}0.707 & \cellcolor[rgb]{0.866666666666667,0.886274509803922,0.513725490196078}0.724 & \cellcolor[rgb]{0.858823529411765,0.882352941176471,0.509803921568627}0.725 & \cellcolor[rgb]{0.847058823529412,0.87843137254902,0.509803921568627}\textbf{0.728} & \cellcolor[rgb]{0.850980392156863,0.87843137254902,0.509803921568627}{0.727} \bigstrut[t]\\
\end{tabular}%
    \end{minipage}
    }
  \label{tab:smeTable}%
\end{table*}%

\subsection{Experimental Setup}
\label{sec:experimentSetup}

In the experiments, we evaluate the power forecasting performance of the data sets (see Section~\ref{sec:dataset}).
As laid out in the CSGE algorithm description of Section~\ref{sec:mmSoftGating}, for each weather forecasting model we perform the forecast using a number of forecasting algorithms.
As optimization for the CSGE technique (see Section~\ref{sec:overallmdl}), a simplex algorithm \citep{Nelder1965} is chosen.
The parameter search is performed in greedy fashion, where the global qualities are optimized followed by the local and the temporal optimization
for both, weather and power forecasting models. The regularization parameter $\zeta$ is chosen in a way that avoids overfitting of the CSGE algorithm to the validation data.

For the evaluation, each data set is split into a training and a test subset. Afterwards, the training set is further split in a $5$-fold cross-validation
into three data sets which are called \emph{parameter set} (3/5), \emph{optimization set} (1/5)  and \emph{validation set} (1/5) for the sake of clarity.
The single power forecasting models for each weather forecasting model are trained using the parameter set. The parameter optimization of the CSGE technique is then performed on the optimization set
and finally evaluated on the validation set. The parameter combination which performed best on the validation set over all folds is chosen as final
model parameterization which is used to compute the final model quality on the test set.

As pointed out, e.g., in  \citep{genslerDeterministic16}, error scores beyond the RMSE are of importance for investigating a forecast.
For our evaluation, we therefore include the RMSE (computed using Eq.~(\ref{eq:rmse})), the coefficient of determination $R^2$, and the skill score for model comparison.
The $R^2 \in [0,1]$ score is the squared correlation coefficient $R$ computed by
\begin{equation}
R^2 = \frac{\big(\sum_{n=1}^{N} (\hat{y}_n - \bar{\hat{y}})({o}_n - \bar{o})\big)^2}{\sum_{n=1}^{N} ({\hat{y}}_n - \bar{\hat{y}})^2 \sum_{n=1}^{N} ({o}_n - \bar{o})^2}.
\end{equation}
It describes the amount of linear correlation between a forecast and the measured values. The optimal value of $R^2$ is $1$ (perfect correlation), whereas $0$ represents no correlation between the forecasts and measurements.
Note that $\bar{\hat{y}}$ is the mean of all issued forecasts in the data set in this case (not an ensemble prediction).
The skill score describes the amount of improvement of an evaluated technique $e_{\mathrm{eval}}$ in comparison to a baseline technique $e_{\mathrm{base}}$. The improvement of the forecasting skill can be computed using the error scores (either RMSE or $R^2$):
\begin{equation}
\label{eq:skillscore}
	\mathrm{Imp} = \frac{e_{\mathrm{base}} - e_{\mathrm{eval}}}{e_{\mathrm{base}}}.
\end{equation}
Furthermore, the number of wins of a particular algorithm is stated.

\begin{table*}[htb]
  \caption{Performance comparison regarding RMSE and the $R^2$ score of a number of multi-model day-ahead ensemble forecasts (see Section \ref{sec:mmResults}). The color coding indicates the quality of each wind farm and power forecasting algorithm from high quality (green) to low quality (red). Bold text highlights the best achieved score for each evaluated data set.}
  \resizebox{\textwidth}{!}{\begin{minipage}{\textwidth}
  \centering
  \scriptsize
  \label{tab:mmeTable2}%
\begin{tabular}{r|rrrrr|rrrrr}
    \cellcolor[rgb]{1,1,1}      & \multicolumn{5}{c|}{RMSE}             & \multicolumn{5}{c}{$R^2$} \\
    \cellcolor[rgb]{1,1,1}      & \cellcolor[rgb]{1,1,1}No Ens. & \cellcolor[rgb]{1,1,1}MME Eq. & \cellcolor[rgb]{1,1,1}MME Fix. & \cellcolor[rgb]{1,1,1}CSGE-S & \cellcolor[rgb]{1,1,1}CSGE-M & \cellcolor[rgb]{1,1,1}No. Ens. & \cellcolor[rgb]{1,1,1}MME Eq. & \cellcolor[rgb]{1,1,1}MME Fix. & \cellcolor[rgb]{1,1,1}CSGE-S & \cellcolor[rgb]{1,1,1}CSGE-M \bigstrut[b]\\
    \hline
    \cellcolor[rgb]{1,1,1}wf1   & \cellcolor[rgb]{0.996078431372549,0.807843137254902,0.498039215686275}0.138 & \cellcolor[rgb]{0.996078431372549,0.788235294117647,0.494117647058824}0.142 & \cellcolor[rgb]{0.996078431372549,0.803921568627451,0.498039215686275}0.139 & \cellcolor[rgb]{0.996078431372549,0.819607843137255,0.498039215686275}0.136 & \cellcolor[rgb]{0.996078431372549,0.823529411764706,0.501960784313725}\textbf{0.135} & \cellcolor[rgb]{0.964705882352941,0.913725490196078,0.517647058823529}0.716 & \cellcolor[rgb]{0.996078431372549,0.913725490196078,0.513725490196078}0.703 & \cellcolor[rgb]{0.972549019607843,0.913725490196078,0.517647058823529}0.714 & \cellcolor[rgb]{0.905882352941176,0.898039215686275,0.513725490196078}0.729 & \cellcolor[rgb]{0.886274509803922,0.890196078431373,0.513725490196078}{\textbf{0.734}} \bigstrut[t]\\
\end{tabular}%
    \end{minipage}
    }
 \end{table*}%
    \begin{table*}[htb]
    \caption{Performance comparison regarding RMSE and the $R^2$ score of a number of multi-model intraday ensemble forecasts (see Section \ref{sec:idResults}). The color coding indicates the quality of each wind farm and power forecasting algorithm from high quality (green) to low quality (red). Bold text highlights the best achieved score for each evaluated data set.}
\resizebox{\textwidth}{!}{\begin{minipage}{\textwidth}
  \centering
  \scriptsize

  \label{tab:idTable}%

    \begin{tabular}{r|rrrrr|rrrrr}
    \cellcolor[rgb]{1,1,1}      & \multicolumn{5}{c|}{RMSE}             & \multicolumn{5}{c}{$R^2$} \\
    \cellcolor[rgb]{1,1,1}      & \cellcolor[rgb]{1,1,1}Pers. & \cellcolor[rgb]{1,1,1}No Ens. & \cellcolor[rgb]{1,1,1}MME Fix. & \cellcolor[rgb]{1,1,1}CSGE-M & \cellcolor[rgb]{1,1,1}CSGE+P & \cellcolor[rgb]{1,1,1}Pers. & \cellcolor[rgb]{1,1,1}No. Ens. & \cellcolor[rgb]{1,1,1}MME Fix. & \cellcolor[rgb]{1,1,1}CSGE-M & \cellcolor[rgb]{1,1,1}CSGE+P \bigstrut[b]\\
    \hline
    \cellcolor[rgb]{1,1,1}wf1   & \cellcolor[rgb]{0.984313725490196,0.619607843137255,0.462745098039216}0.218 & \cellcolor[rgb]{1,0.898039215686275,0.513725490196078}0.122 & \cellcolor[rgb]{1,0.87843137254902,0.509803921568627}0.129 & \cellcolor[rgb]{1,0.913725490196078,0.517647058823529}0.118 & \cellcolor[rgb]{0.980392156862745,0.913725490196078,0.513725490196078}\textbf{0.112} & \cellcolor[rgb]{0.980392156862745,0.623529411764706,0.458823529411765}0.404 & \cellcolor[rgb]{0.874509803921569,0.886274509803922,0.513725490196078}0.773 & \cellcolor[rgb]{0.945098039215686,0.905882352941176,0.517647058823529}0.756 & \cellcolor[rgb]{0.780392156862745,0.858823529411765,0.505882352941176}0.797 & \cellcolor[rgb]{0.705882352941177,0.83921568627451,0.501960784313725}\textbf{0.816} \bigstrut[t]\\
\end{tabular}%
  \end{minipage}
  }
\end{table*}%

\subsection{Day-Ahead Performance on Single Weather Forecasting Model}
\label{sec:smResults}

This experiment aims at comparing the performance of the CSGE technique to other techniques using a single weather forecasting model in comparison to other forecasting algorithms and ensembles on the day-ahead forecasting horizon
($k_{\mathrm{min}} = 24$, $k_{\mathrm{max}}=48$, $\Delta = 1 h$)
on the
data set of Section \ref{sec:dataset}.
As there is only one weather forecasting model, the number of weighting dimensions is reduced to the $3$ power forecasting model based weighting factors
$w^{(\varphi|\psi)} = w^{(\varphi|\psi)}_g \cdot w^{(\varphi|\psi)}_l \cdot w^{(\varphi|\psi)}_{k}$ for each power forecasting model.
As power forecasting techniques, we include some state of the art model types, namely feed-forward artificial neural networks (ANN), linear regression (Lin.~Reg.),
a boosting (LSBoost) \citep{Schapire1990} and a bagging (Bag.) \citep{Breiman1996} ensemble forecasting technique (with decision trees), and the ensemble technique proposed in \citep{Gensler2016ensemble} (CE),
which uses a rudimentary form of coopetitive soft-gating.
We state the RMSE, as well as the $R^2$ score, the respective forecasting skill factor (with linear regression as baseline), and the number of wins.
We use Matlab implementations for all algorithms.

As can be seen from Table \ref{tab:smeTable}, the wind farm data sets have a varying error regarding the forecasting accuracy. Some wind farms are well predictable regarding the RMSE (e.g., wf3, wf9, wf13, wf41), whereas the forecasting algorithms
struggle with other wind farms (e.g., wf6, wf20, wf31), as can be derived from the respective scores and from the color coding in the table.
Regarding the overall performance, all evaluated ensemble techniques perform better on average in comparison to the forecasting techniques based on a single forecasting algorithm, as can be seen from the forecasting skill (next to last row).
The best ensemble techniques on the data set are the CSGE variants, where the suffixes in the table indicate locality assessment using either the nearest neighbor (CSGE-K), or interpolation technique CSGE-I from \citep{Gensler2016ensemble}. The CSGE variants are closely followed by the CE and the bagging technique. The boosting algorithm performs weaker than the other ensemble algorithms.
The $R^2$ score gives similar indication as the RMSE score. While there are some differences regarding single data sets, regarding the average $R^2$ score,
the overall order of the best performing algorithms remains the same.
It must be mentioned that in this scenario, the CSGE techniques are not exploited to their full extend, as there is only a single weather forecasting model to make use of. Therefore, only $3$ of the $6$ weighting parameters are applied.
In the following experiment (Section~\ref{sec:mmResults}), the performance using multiple weather forecasting models is evaluated, fully exploiting the advantages of the CSGE technique.

\subsection{Day-Ahead Performance on Multiple Weather Forecasting Models}
\label{sec:mmResults}

This experiment aims at comparing the performance of the CSGE technique to other Multi-Model-Ensemble (MME) techniques.
In the experiment, we use $3$ day-ahead weather forecasting models. The three models are available for $36$ windfarms for a time-period of $22$ months.
The time periods do not overlap entirely with the data used in Section~\ref{sec:smResults}.

For this comparison, we choose the following techniques:
\begin{enumerate}
\itemsep0em 
  \item No Ens.: The power forecasting model is the overall best non-ensemble forecasting technique (i.e., an ANN model) with the single best weather forecasting model (determined using the table in Fig.~\ref{fig:mmAll}).
  \item MME Eq.: The model is computed using the best non-ensemble forecasting technique (ANN model) with all weather forecasting models. The model forecasts are averaged, i.e., $w^{(j)} = \frac{1}{J}$. This technique, is, e.g., utilized in \citep{Hagedorn2005}.
  \item MME Fix.: The model is computed using the best non-ensemble forecasting technique (ANN model) with all weather forecasting models. The models are weighted with respect to their global quality using Eq.~(\ref{eq:wmglobWeighting}) with $\eta=2$,
  such that the best performing weather forecasting models get the highest impact in the weighting. This technique is employed, for instance, in \citep{Rajagopalan2002}.
  \item CSGE-S: The CSGE technique is applied using a \emph{single} power forecasting model for each of the three weather forecasting models, the best non-ensemble forecasting technique (ANN model). The CSGE locality assessment is performed using the nearest neighbor technique.
  \item CSGE-M: The CSGE technique using \emph{multiple} power forecasting models for each weather forecasting model is applied. The CSGE locality assessment is performed using the nearest neighbor technique.
\end{enumerate}

The results of the day-ahead MME are shown in Table \ref{tab:mmeTable2}.
As can be seen from this table, on average, all ensemble models perform better than a model which only uses a single weather forecasting model.
Therein, the CSGE technique using multiple power forecasting models (CSGE-M) performs best regarding the RMSE score, followed by the CSGE-S, the
MME Fix., and the MME Eq.~models. For $R^2$, both CSGE variants perform comparably well, followed by the other techniques.
For the data sets, the CSGE-M performs $8.78$\% better than the best non-ensemble technique using the single best weather forecasting model.
The other multi-model ensemble techniques also yield better results than the baseline technique.
While there are some data sets with low RMSE (e.g., wf6, wf9), this error is due to a low value of the target predictands in these data sets,
and not due to a good predictability regarding correlation, as can be seen from the $R^2$ score.
Particular attention should be drawn to {wf9}, where the CSGE-M technique has a low $R^2$ score.
This may be due to an overfitting of the ensemble model, if the regularization parameter $\zeta$ is chosen too low. Still,
the CSGE-M technique exceeds the performance of the other approaches regarding the RMSE on this data set.

As can be seen from the performance of the examined ensemble techniques, the use of multiple weather forecasting models clearly improves the performance, e.g., in MME Eq.
The use of a global weighting (such as conducted in MME Fix.) further increases the performance.
The use of local and temporal weighting in addition to the global weighting (as performed in CSGE-S) increases the performance, even if just one power forecasting model is used.
The scores can further be improved when performing the weighting for both, weather and power forecasting models (as performed in CSGE-M).

\begin{figure*}[tb]
\centering
    \subfloat[Day-ahead forecast when adding increasingly well performing weather forecasting models.]{%
	\includegraphics[width=0.23\textwidth]{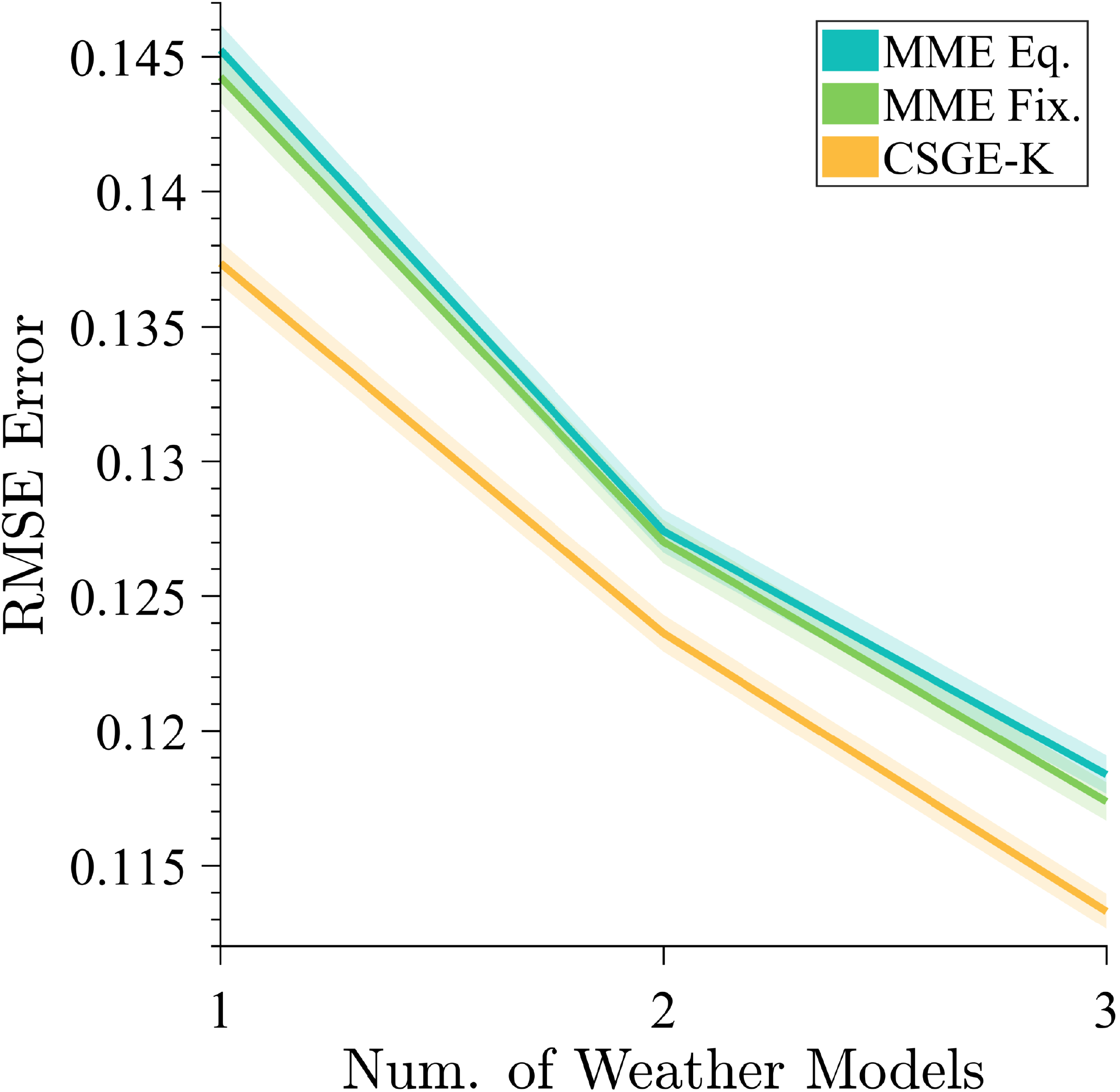}
	\label{fig:mmeBestLast}%
	}
\hfil
	\subfloat[Day-ahead forecast when adding increasingly weak performing weather forecasting models.]{%
	\includegraphics[width=0.23\textwidth]{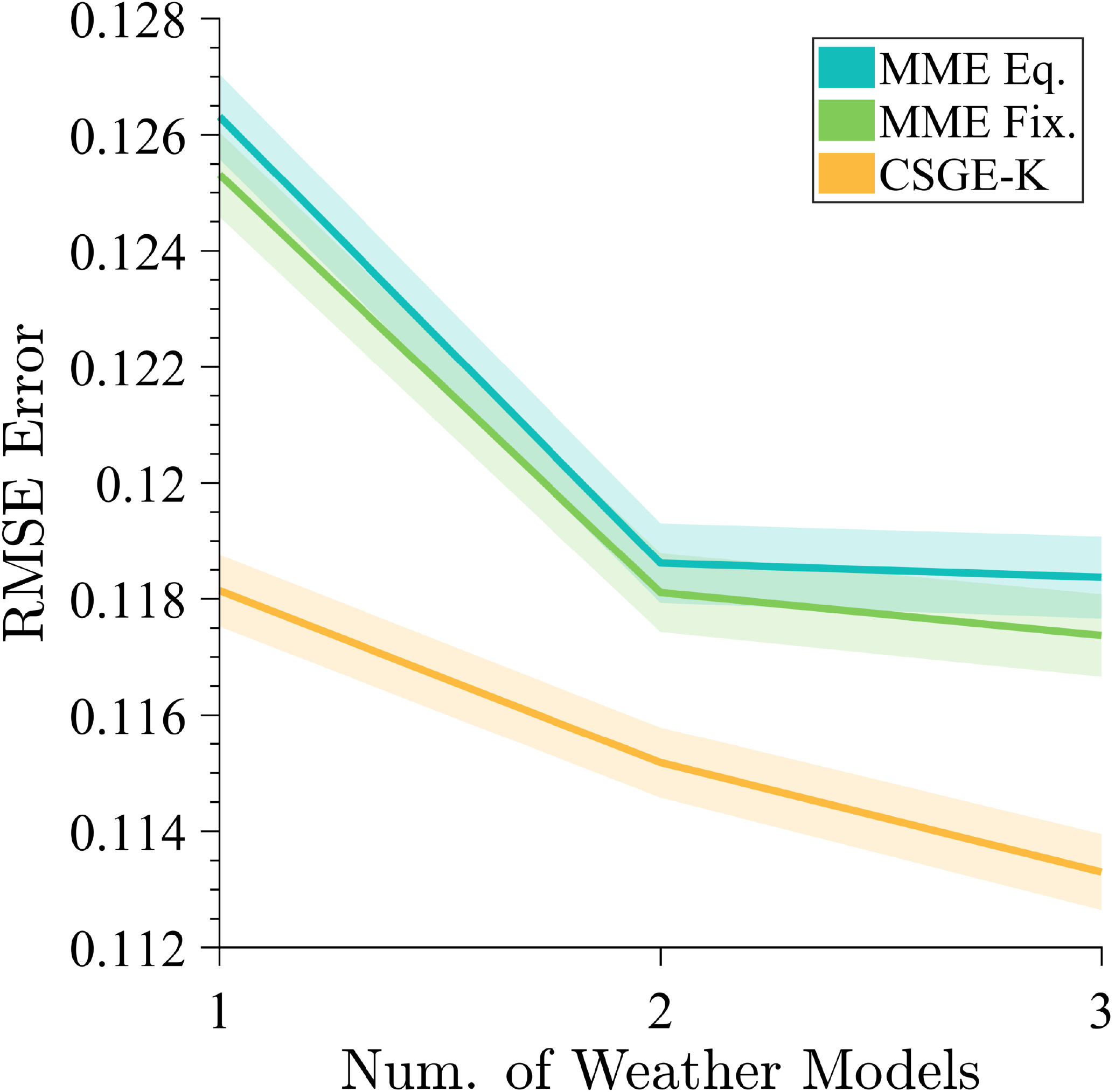}
	\label{fig:mmeBestFirst}%
	}
	\hfil
	\subfloat[Intraday forecast when adding increasingly well performing weather forecasting models.]{%
	\includegraphics[width=0.23\textwidth]{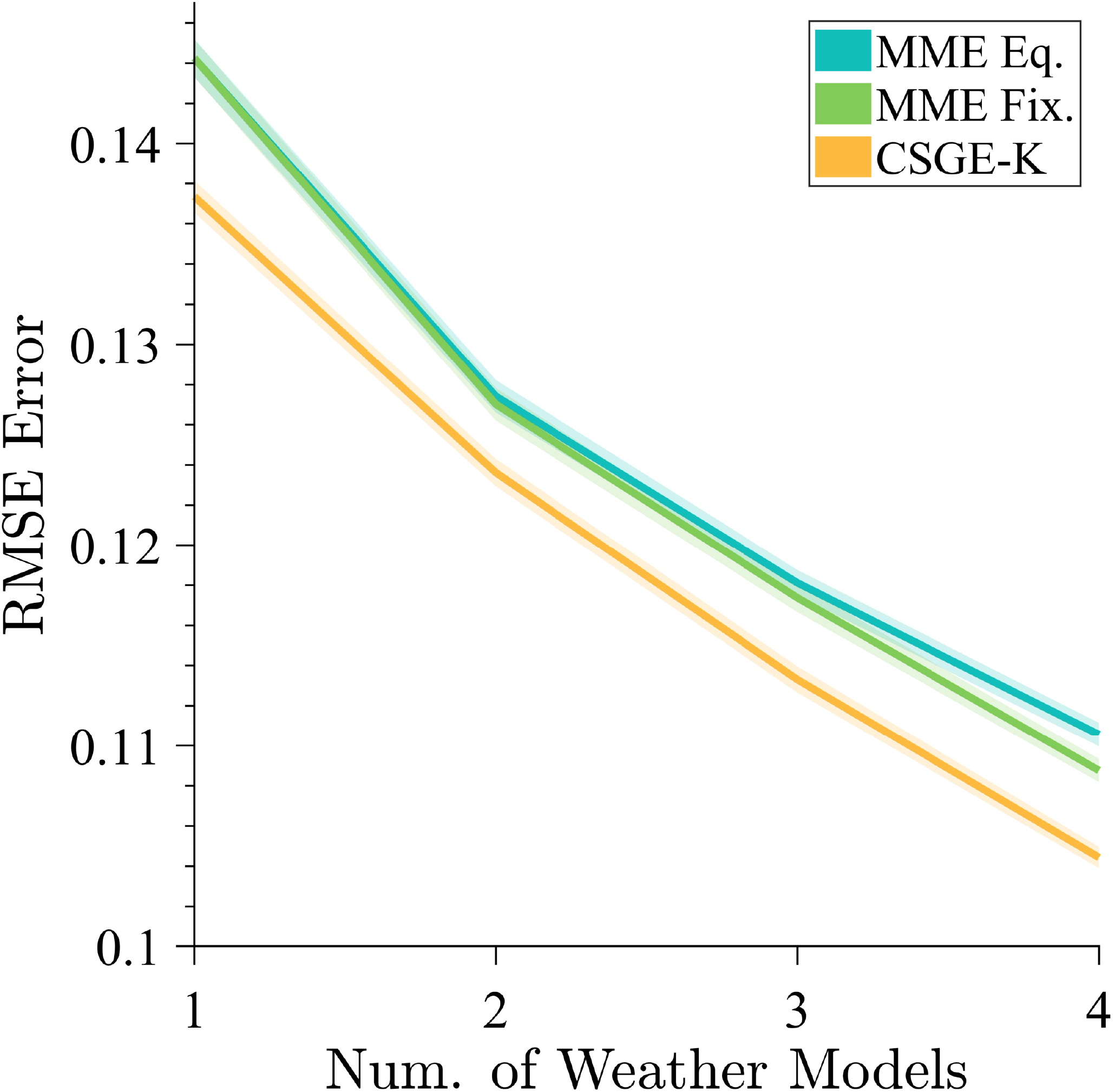}
	\label{fig:mmeIDBestLast}%
	}
\hfil
\subfloat[Intraday forecast when adding increasingly weak performing weather forecasting models.]{%
	\includegraphics[width=0.23\textwidth]{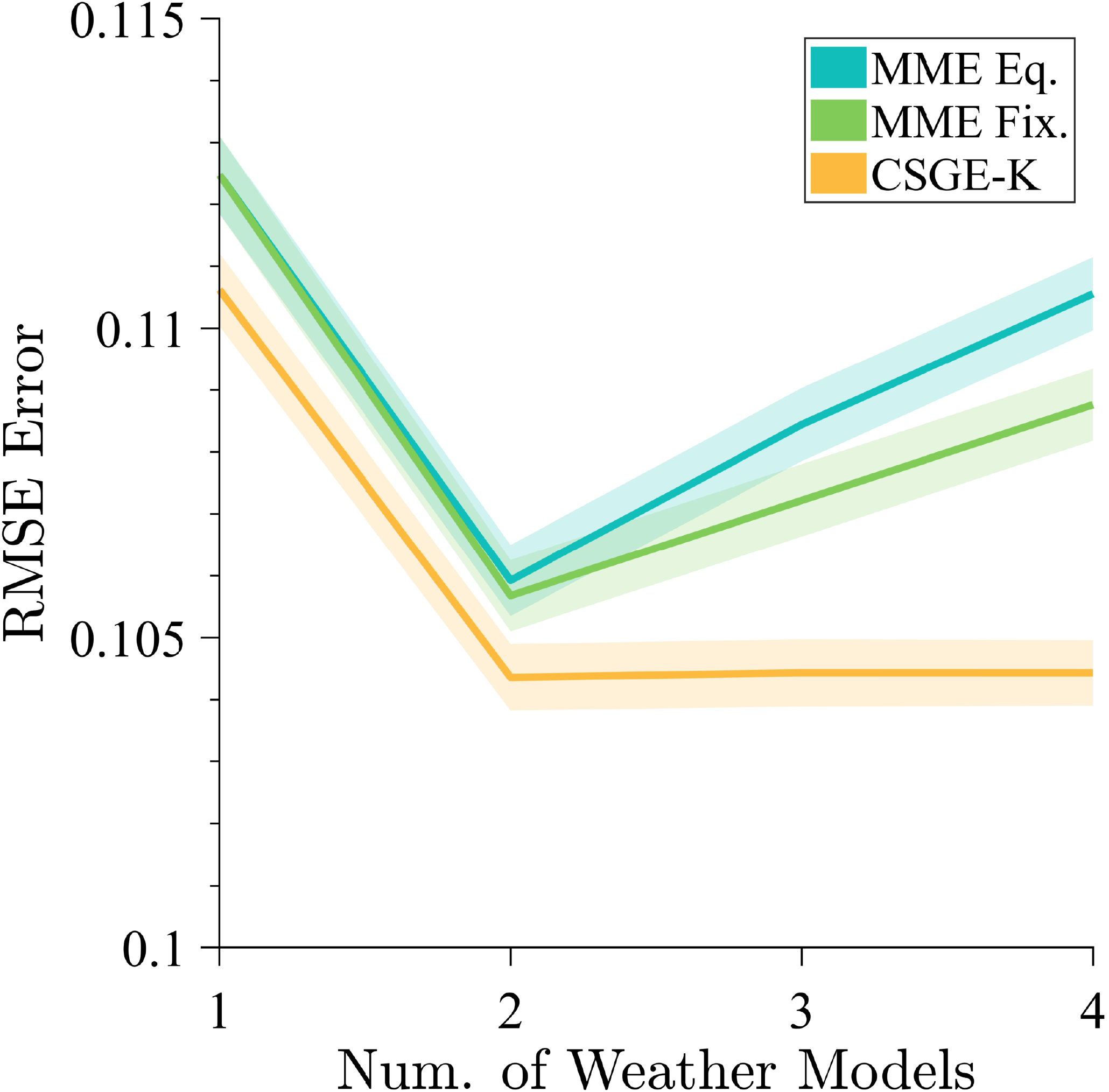}
	\label{fig:mmeIDBestFirst}%
	}
	\caption{Average development of the RMSE quality of a number of ensemble techniques when including additional weather forecasting models to the ensemble for both, day-ahead and intraday forecasts, using the EuropeWindFarm data set collection. When adding increasingly well performing weather forecasting models, all
ensemble algorithms benefit from the inclusion of additional weather forecasting models (Figs.~\ref{fig:mmeBestLast} and \ref{fig:mmeIDBestLast}). When adding increasingly weak performing models to the ensemble, the CSGE-K technique is able to better use this additional information.
The comparison techniques' performance, on the other hand, remains constant (see Fig.~\ref{fig:mmeBestFirst}), or even decreases (see Fig.~\ref{fig:mmeIDBestFirst}).}
	\label{fig:mmeDevelopment}%
\end{figure*}

\subsection{Intraday Performance as Lagged Multi-Model Ensemble}
\label{sec:idResults}

This experiment evaluates the ability of the proposed method to work as a lagged multi-model ensemble. Therefore, an intraday weather forecasting model is added to the three existing weather forecasting models,
and an \emph{intraday} forecast is performed ($k_{\mathrm{min}} = 1$, $k_{\mathrm{max}}=24$, $\Delta = 1 h$). The overall forecast therefore is created from one intraday forecast and three day-ahead forecasts.
As comparison techniques, we include the pure persistence method, the best non-ensemble technique (No Ens., which is an ANN), the best conventional multi-model ensemble technique MME Fix.~from the previous section, and the proposed algorithm, without (CSGE-K) and additionally including the persistence method in the CSGE-K approach (CSGE+P).

The results of the experiment are shown in Table~\ref{tab:idTable}. As expected, the persistence method performs weakest. All models yield better
forecasting results due to the new intraday model which is included in the forecast. Within the ensemble techniques, the proposed CSGE-K technique exceeds the performance of the
MME fixed technique, whereas the inclusion of the persistence method as ensemble member (CSGE+P) yields further quality improvements regarding both, RMSE and $R^2$ score.
The CSGE+P technique exceeds the persistence models performance by $49.13\%$ regarding RMSE, followed by the CSGE-K technique ($48.24\%$), and the fixed ensemble.
The $R^2$ scores show very similar results.

\subsection{Performance Development Using Varying Number of Weather Forecasting Models}
\label{sec:mmGains}

This experiment gives insight into a practical problem in power forecasting: Given a number of weather forecasting models available, which of those models should be included, and will the overall quality eventually even be lowered when
worse performing models or models with unknown performance are added to the overall ensemble.
This section therefore describes the dependence of the performance of the techniques in Sections \ref{sec:mmResults} and \ref{sec:idResults} when adding additional weather forecasting models to the forecast.
In order to evaluate the performance, the average RMSE of the algorithms used in Sections \ref{sec:mmResults} and \ref{sec:idResults} are computed with a varying number of weather forecasting models.
Figs. \ref{fig:mmeBestLast} and \ref{fig:mmeIDBestLast} show the dependence of the performance when the weakest model is chosen as first model, and increasingly
good weather forecasting models are added. The bounds indicate the model variance on the data sets.
Figs. \ref{fig:mmeBestFirst} and \ref{fig:mmeIDBestFirst}, on the other hand, show the development of the performance when the best weather forecasting model is chosen first and worse models are added subsequently.

For the day-ahead variants of Figs \ref{fig:mmeBestLast}, and \ref{fig:mmeBestFirst}, all ensemble models benefit from increasing the number of weather forecasting models, even if the additional models perform worse in comparison to the first weather forecasting model.
Fig.~\ref{fig:mmeBestLast} shows the RMSE error when including increasingly good performing models. As can be expected, the performance is clearly higher when adding additional weather forecasting models.
As can be seen from Fig.~\ref{fig:mmeBestFirst}, the performance improvement drops when including a third weather forecasting model and the added model performs worse than the already included models, in this case. The approaches for comparison benefit more when including a second model.
However, the CSGE-K is able to benefit more from the inclusion of a third weather forecasting model. In particular the MME Eq.~technique is not able to regulate the model influence precisely enough and therefore
barely benefits from the inclusion of a third model.

For intraday forecasts, the behavior is similar to the day-ahead forecast when adding increasingly well performing models, as is indicated in Fig \ref{fig:mmeIDBestLast}. When including an intraday model as fourth model
in addition to the three day-ahead forecasts, all models benefit about equally from the inclusion of the fourth model.
An interesting phenomenon is visible in Fig.~\ref{fig:mmeIDBestFirst}, i.e., when adding additional day-ahead weather forecasting models to an intraday weather forecasting model.
All models benefit from the inclusion of the best-performing day-ahead weather forecasting model. The maximum benefit of the inclusion of multiple weather forecasting models seems to be reached at this point, as no model is able to benefit from additional
weather forecasting models.
However, when including further weather forecasting models, the model performance of the comparison approaches \emph{decreases}, i.e., the inclusion of additional models irritates the ensemble forecast.
The CSGE-K technique, on the other hand, is able to recognize the weak performing weather forecasting models and is able to reduce the respective weights in a way that does not negatively affect the overall model performance.

\subsection{Discussion of Applicability}
\label{sec:discussion}

Beside the performance of the proposed approach, there are some other model properties worth mentioning.

\subsubsection*{Failure Mode}
  The ensemble weights $w^{(j)}$ are computed dynamically for each forecasting time step for which a forecast is performed.
  The CSGE algorithm computes the weights for the respective weighting categories using all available forecasts \emph{at that particular time}.
  Thereby, the ensemble can create a forecast even if some power forecasting models fail to create a prediction, or if
  the NWP forecast for a particular weather forecasting model fails to be delivered.
  The proposed technique is then able to retain the optimal weighting performance given the circumstances.
  \subsubsection*{Ensemble of Opportunity }
  \label{sec:ensOpportunity} Given this ``failure mode'' of the ensemble, an applied forecasting system can then also be designed to
  work as an ensemble of opportunity. A number of power forecasting models and weather forecasting models can be prepared (pretrained) for application,
  however, not all power forecasting models (and/or not all weather forecasting models) have to be evaluated every time a prediction is created.
  The number of evaluated power forecasting models can be chosen dynamically.
  This number could depend on one or more of the following aspects:
  \begin{itemize}[leftmargin=*]
  \itemsep0em 
    \item Estimated difficulty of the current forecasting task, e.g., derived from weather situation.
    \item Uncertainty of the current prediction in a probabilistic forecast, e.g., from inner-ensemble disagreement.
    \item Expected quality $w^{(j)}$ in the ensemble, i.e., elimination of models with low weight if too high costs are present.
    \item Criticality / importance of a precise forecast to the power grid operation.
    \item Necessity of fast delivery time.
    \item Financial cost of a reported deviation from the true power generation.
    \item Available computation capacity in a computing cluster.
  \end{itemize}
  \subsubsection*{Cost/Reward Functions}
  \label{sec:costReward} Given the availability of (hybrid) cloud computing solutions and e.g., computing concepts such as cheap preemptible virtual machines, the employment
  of using cost/reward functions for financial optimizations can make sense, given the trade-off of investment in computing costs and gain in quality. The above-mentioned factors can have an impact on the design of the cost/reward function.
\subsubsection*{Parameter Determination} The proposed ensemble technique has a low number of parameters only. Besides the power forecasting models to choose, the main parameter is
the regularization parameter $\zeta$ which regulates the amount of coopetitive soft gating.
While choosing an improper value of $\zeta$ negatively affects the model performance, it still will perform either as a static model averaging (too much regularization) or
pure gating, possibly with overfitting (too little regularization).
\subsubsection*{Variants of CSGE}
In the article, a nearest neighbor technique to assess the local (weather-dependent) quality is proposed (in Section~\ref{sec:loc}).
However, one can imagine other possibilities to assess the local quality, such as a multi-linear interpolation (MLI) technique, which is presented in \citep{Gensler2016ensemble}.
As shown in the experiments in the referenced article, the two techniques behave very similar regarding forecasting performance. The respective technique
should therefore be chosen depending on the size of the historic training data set. The MLI has a training phase, however, it does then no longer need the training data in order
to operate. This approach is, therefore, well-suited for large data sets. The CSGE-K technique is able to work without model training, however, it does have to
search the data set on every query. The technique is, therefore, better suited for smaller data sets.
\subsubsection*{Parameter Optimization}
An advantage of the proposed CSGE technique is that the ensemble training (the determination of the weights $\bm{\eta}$) is a post-processing step of the training of the ensemble members. The ensemble is trained by Eq.~\ref{eq:optimization}
which optimizes the weights rather than the single ensemble member forecasts.
This, in turn, means that during ensemble training, the forecasts of the ensemble members do \emph{not} have
to be recomputed when varying $\bm{\eta}$. The evaluation of the model fitness therefore is swiftly possible.
Furthermore, the weights gradually change when varying the coopetitive soft gating strengths $\bm{\eta}$, thus the optimization problem is smooth.
In this article, the parameter optimization was performed in a greedy fashion for the sake of simplicity and speed. However, one can also think of overall optimization using
techniques such as simulated annealing, stochastic gradient descent, or particle swarms, possibly leading to an even better optimization.
\subsubsection*{On-Line Weight Improvement}
The proposed technique can be extended to update the weighting methods consistently ``on-line'' when observing novel model input data by permanently learning on novel observations. The respective power measurements have to be included in this setup in order
to enable a meaningful feedback.

\section{Conclusion and Outlook}
\label{sec:conclusion}

In this article, we proposed a novel multi-scheme ensemble based on a technique we call coopetitive soft gating.
The technique aims to adaptively exploit the strengths of various power forecasting and weather forecasting models
regarding their overall global quality, their lead time dependent quality, and their weather situation-dependent local quality.
The technique is able to yield superior results in comparison to other forecasting algorithms and ensemble techniques as has been shown in a number of experiments on publicly available data sets.
The flexible structure of the ensemble technique enables an employment of the proposed algorithm
for a number of lead times and using a flexible number of power and weather forecasting models, respectively.

In the future, we aim to implement some of the schemes to enable the ensemble technique to work as an ensemble of opportunity using
some of the influencing factors described in Section \ref{sec:ensOpportunity}.
We also aim to develop a technique for assessing the uncertainty of a forecast depending on the weights computed using coopetitive soft gating.
Therein, our goal is to develop an adaptive model pruning method depending on the particular situation to forecast when the expected
quality of a power forecasting model for a particular forecasting situation (e.g., weather situation or forecasting time step) is low to avoid unnecessary computations with marginal impact.
Also, the inclusion of more sophisticated base predictors, such as (deep) neural network structures (e.g., analyzed in~\cite{Raabe2016}), in the ensemble may be of interest.


\section*{Acknowledgment}

This article results from the project BigEnergy (HA project no.~472/15-14), which is funded
in the framework of Hessen ModellProjekte,
financed with funds of the LOEWE – Landes-Offensive
zur Entwicklung Wissenschaftlich-\"okonomischer
Exzellenz, F\"orderlinie 3: KMU-Verbundvorhaben
(State Offensive for the Development of Scientific
and Economic Excellence).
We want to thank the Enercast GmbH
for providing the data sets.




\bibliographystyle{elsarticle-harv}
\bibliography{library}

\begin{thebibliography}{50}
\expandafter\ifx\csname natexlab\endcsname\relax\def\natexlab#1{#1}\fi
\expandafter\ifx\csname url\endcsname\relax
  \def\url#1{\texttt{#1}}\fi
\expandafter\ifx\csname urlprefix\endcsname\relax\def\urlprefix{URL }\fi

\bibitem[{Alessandrini et~al.(2013)Alessandrini, Sperati, and
  Pinson}]{Alessandrini2013}
Alessandrini, S., Sperati, S., Pinson, P., 2013. {A comparison between the
  ECMWF and COSMO Ensemble Prediction Systems applied to short-term wind power
  forecasting on real data}. Applied Energy 107, 271--280.

\bibitem[{Bessa et~al.(2012)Bessa, Miranda, Botterud, Zhou, and
  Wang}]{Bessa2012a}
Bessa, R.~J., Miranda, V., Botterud, A., Zhou, Z., Wang, J., 2012.
  {Time-adaptive quantile-copula for wind power probabilistic forecasting}.
  Renewable Energy 40~(1), 29--39.

\bibitem[{Breiman(1996)}]{Breiman1996}
Breiman, L., 1996. {Bagging Predictors}. Machine Learning 24~(421), 123--140.

\bibitem[{Breiman(2001)}]{Breiman2001}
Breiman, L., 2001. {Random forests}. Machine Learning 45~(1), 5--32.

\bibitem[{Costa et~al.(2008)Costa, Crespo, Navarro, Lizcano, Madsen, and
  Feitosa}]{Costa2008}
Costa, A., Crespo, A., Navarro, J., Lizcano, G., Madsen, H., Feitosa, E., 2008.
  {A review on the young history of the wind power short-term prediction}.
  Renewable and Sustainable Energy Reviews 12~(6), 1725--1744.

\bibitem[{Craig et~al.(2002)Craig, Gadgil, and Koomey}]{Craig2002}
Craig, P.~P., Gadgil, A., Koomey, J.~G., 2002. {What can history teach us? A
  Retrospective Examination of Long-Term Energy Forecasts for the United
  States}. Annual Review of Energy and the Environment 27~(1), 83--118.

\bibitem[{{Delle Monache} et~al.(2013){Delle Monache}, Eckel, Rife, Nagarajan,
  and Searight}]{DelleMonache2013}
{Delle Monache}, L., Eckel, F.~A., Rife, D.~L., Nagarajan, B., Searight, K.,
  2013. {Probabilistic Weather Prediction with an Analog Ensemble}. Monthly
  Weather Review 141~(10), 3498--3516.

\bibitem[{Dietterich(2000)}]{Dietterich2000}
Dietterich, T.~G., 2000. {Ensemble Methods in Machine Learning}. Multiple
  Classifier Systems 1857, 1--15.

\bibitem[{Duan et~al.(2007)Duan, Ajami, Gao, and Sorooshian}]{Duan2007}
Duan, Q., Ajami, N.~K., Gao, X., Sorooshian, S., 2007. {Multi-model ensemble
  hydrologic prediction using Bayesian model averaging}. Advances in Water
  Resources 30~(5), 1371--1386.

\bibitem[{Foley et~al.(2012)Foley, Leahy, Marvuglia, and McKeogh}]{Foley2012}
Foley, A.~M., Leahy, P.~G., Marvuglia, A., McKeogh, E.~J., 2012. {Current
  methods and advances in forecasting of wind power generation}. Renewable
  Energy 37~(1), 1--8.

\bibitem[{Gensler(2016)}]{Gensler2016}
Gensler, A., 2016. {EuropeWindFarm Data Set, last accessed 2018-03-15}.
\newline\urlprefix\url{http://ies-research.de/Software}

\bibitem[{Gensler et~al.(2016{\natexlab{a}})Gensler, Henze, Sick, and
  Raabe}]{Raabe2016}
Gensler, A., Henze, J., Sick, B., Raabe, N., 2016{\natexlab{a}}. {Deep Learning
  for Solar Power Forecasting – An Approach Using Autoencoder and LSTM Neural
  Networks}. In: Proceedings of the IEEE International Conference on Systems,
  Man, and Cybernetics (SMC16). Budapest, Hungary, pp. 2858--2865.

\bibitem[{Gensler and Sick(2016)}]{Gensler2016ensemble}
Gensler, A., Sick, B., 2016. {Forecasting wind power -- an ensemble technique
  with gradual coopetitive weighting based on weather situation}. In:
  Proceedings of the IEEE International Joint Conference on Neural Networks
  (IJCNN16). Vancouver, Canada, pp. 4976--4984.

\bibitem[{Gensler and Sick(2017)}]{genslerpcsge17}
Gensler, A., Sick, B., 2017. {Probabilistic Wind Power Forecasting: A
  Multi-Scheme Ensemble Technique With Gradual Coopetitive Soft Gating}. In:
  Proceedings of the 9th IEEE Symposium Series on Computational Intelligence
  (SSCI17). Honolulu, USA, pp. 1803--1812.

\bibitem[{Gensler et~al.(2016{\natexlab{b}})Gensler, Sick, and
  Pankraz}]{Gensler2016Analog}
Gensler, A., Sick, B., Pankraz, V., 2016{\natexlab{b}}. {An analog
  ensemble-based similarity search technique for solar power forecasting}. In:
  Proceedings of the IEEE International Conference on Systems, Man, and
  Cybernetics (SMC16). Budapest, Hungary, pp. 2850--2857.

\bibitem[{Gensler et~al.(2016{\natexlab{c}})Gensler, Sick, and
  Vogt}]{genslerDeterministic16}
Gensler, A., Sick, B., Vogt, S., 2016{\natexlab{c}}. {A review of deterministic
  error scores and normalization techniques for power forecasting algorithms}.
  In: Proceedings of the 8th IEEE Symposium Series on Computational
  Intelligence (SSCI16). Athens, Greece, pp. 1--9.

\bibitem[{Giebel et~al.(2011)Giebel, Brownsword, Kariniotakis, Denhard, and
  Draxl}]{Giebel2011}
Giebel, G., Brownsword, R., Kariniotakis, G., Denhard, M., Draxl, C., 2011.
  {The State-Of-The-Art in Short-Term Prediction of Wind Power: A Literature
  Overview, 2nd Edition}. Tech. rep., Lyngby, Denmark.

\bibitem[{G{\"{o}}nen and Alpaydın(2011)}]{Gonen2011}
G{\"{o}}nen, M., Alpaydın, E., 2011. {Multiple Kernel Learning Algorithms}.
  Journal of Machine Learning Research 12, 2211--2268.

\bibitem[{Hagedorn et~al.(2005)Hagedorn, Doblas-Reyes, and
  Palmer}]{Hagedorn2005}
Hagedorn, R., Doblas-Reyes, F.~J., Palmer, T.~N., 2005. {The rationale behind
  the success of multi-model ensembles in seasonal forecasting - I. Basic
  concept}. Tellus, Series A: Dynamic Meteorology and Oceanography 57~(3),
  219--233.

\bibitem[{Ho(1998)}]{Ho1998}
Ho, T.~K., 1998. {The random subspace method for constructing decision
  forests}. IEEE Transactions on Pattern Analysis and Machine Intelligence
  20~(8), 832--844.

\bibitem[{Holttinen et~al.(2012)Holttinen, Milligan, Ela, Menemenlis,
  Dobschinski, Rawn, Bessa, Flynn, G{\'{o}}mez-L{\'{a}}zaro, and
  Detlefsen}]{Holttinen2012}
Holttinen, H., Milligan, M., Ela, E., Menemenlis, N., Dobschinski, J., Rawn,
  B., Bessa, R.~J., Flynn, D., G{\'{o}}mez-L{\'{a}}zaro, E., Detlefsen, N.~K.,
  2012. {Methodologies to determine operating reserves due to increased wind
  power}. IEEE Transactions on Sustainable Energy 3~(4), 713--723.

\bibitem[{Hong et~al.(2014)Hong, Wilson, and Xie}]{Hong2014b}
Hong, T., Wilson, J., Xie, J., 2014. {Long term probabilistic load forecasting
  and normalization with hourly information}. IEEE Transactions on Smart Grid
  5~(1), 456--462.

\bibitem[{Kohavi and Wolpert(1996)}]{Kohavi1996}
Kohavi, R., Wolpert, D.~H., 1996. {Bias plus variance decomposition for
  zero-one loss functions}. Proceedings of the 13th International Conference on
  Machine Learning (ICML96), 275--283.

\bibitem[{Lei et~al.(2009)Lei, Shiyan, Chuanwen, Hongling, and Yan}]{Lei2009}
Lei, M., Shiyan, L., Chuanwen, J., Hongling, L., Yan, Z., 2009. {A review on
  the forecasting of wind speed and generated power}. Renewable and Sustainable
  Energy Reviews 13~(4), 915--920.

\bibitem[{Li et~al.(2011)Li, Shi, and Zhou}]{Li2011}
Li, G., Shi, J., Zhou, J., 2011. {Bayesian adaptive combination of short-term
  wind speed forecasts from neural network models}. Renewable Energy 36~(1),
  352--359.

\bibitem[{Loebecke et~al.(1999)Loebecke, {Van Fenema}, and
  Powell}]{Loebecke1999}
Loebecke, C., {Van Fenema}, P.~C., Powell, P., 1999. {Co-opetition and
  knowledge transfer}. Database for Advances in Information Systems 30~(2),
  14--25.

\bibitem[{Matos and Bessa(2011)}]{Matos2011}
Matos, M.~A., Bessa, R.~J., 2011. {Setting the operating reserve using
  probabilistic wind power forecasts}. IEEE Transactions on Power Systems
  26~(2), 594--603.

\bibitem[{Mendes-Moreira et~al.(2012)Mendes-Moreira, Soares, Jorge, and
  Sousa}]{Mendes-Moreira2012}
Mendes-Moreira, J., Soares, C., Jorge, A.~M., Sousa, J. F.~D., 2012. {Ensemble
  approaches for regression}. ACM Computing Surveys 45~(1), 1--40.

\bibitem[{Mirasgedis et~al.(2006)Mirasgedis, Sarafidis, Georgopoulou, Lalas,
  Moschovits, Karagiannis, and Papakonstantinou}]{MIRASGEDIS2006}
Mirasgedis, S., Sarafidis, Y., Georgopoulou, E., Lalas, D., Moschovits, M.,
  Karagiannis, F., Papakonstantinou, D., 2006. {Models for mid-term electricity
  demand forecasting incorporating weather influences}. Energy 31~(2-3),
  208--227.

\bibitem[{Mittermaier(2007)}]{Mittermaier2007}
Mittermaier, M.~P., 2007. {Improving short-range high-resolution model
  precipitation forecast skill using time-lagged ensembles}. Quarterly Journal
  of the Royal Meteorological Society 133~(October), 1487--1500.

\bibitem[{Nelder and Mead(1965)}]{Nelder1965}
Nelder, J., Mead, R., 1965. {A simplex method for function minimization}.
  Computer Journal 7~(4), 308--313.

\bibitem[{Pierro et~al.(2016)Pierro, Bucci, {De Felice}, Maggioni, Moser,
  Perotto, Spada, and Cornaro}]{Pierro2016}
Pierro, M., Bucci, F., {De Felice}, M., Maggioni, E., Moser, D., Perotto, A.,
  Spada, F., Cornaro, C., 2016. {Multi-Model Ensemble for day ahead prediction
  of photovoltaic power generation}. Solar Energy 134, 132--146.

\bibitem[{Pinson and Girard(2012)}]{Pinson2012}
Pinson, P., Girard, R., 2012. {Evaluating the quality of scenarios of
  short-term wind power generation}. Applied Energy 96, 12--20.

\bibitem[{Pinson and Kariniotakis(2004)}]{Pinson2004}
Pinson, P., Kariniotakis, G., 2004. {On-line assessment of prediction risk for
  wind power production forecasts}. Wind Energy 7~(2), 119--132.

\bibitem[{Pinson et~al.(2009)Pinson, Nielsen, Madsen, and
  Kariniotakis}]{Pinson2009d}
Pinson, P., Nielsen, H., Madsen, H., Kariniotakis, G., 2009. {Skill forecasting
  from ensemble predictions of wind power}. Applied Energy 86~(7-8),
  1326--1334.

\bibitem[{Platt(1999)}]{Platt1999}
Platt, J., 1999. {Probabilistic outputs for support vector machines and
  comparisons to regularized likelihood methods}. Advances in large margin
  classifiers 10~(3), 61--74.

\bibitem[{Rajagopalan et~al.(2002)Rajagopalan, Lall, and
  Zebiak}]{Rajagopalan2002}
Rajagopalan, B., Lall, U., Zebiak, S.~E., 2002. {Categorical Climate Forecasts
  through Regularization and Optimal Combination of Multiple GCM Ensembles}.
  Monthly Weather Review 130~(7), 1792--1811.

\bibitem[{Ren et~al.(2015)Ren, Suganthan, and Srikanth}]{Ren2015}
Ren, Y., Suganthan, P., Srikanth, N., 2015. {Ensemble methods for wind and
  solar power forecasting - A state-of-the-art review}. Renewable and
  Sustainable Energy Reviews 50, 82--91.

\bibitem[{Ren et~al.(2016)Ren, Zhang, and Suganthan}]{Ren2016}
Ren, Y., Zhang, L., Suganthan, P.~N., 2016. {Ensemble Classification and
  Regression— Recent Developments, Applications and Future Directions}. IEEE
  Computational Intelligence Magazine 11~(1), 41--53.

\bibitem[{Schapire(1990)}]{Schapire1990}
Schapire, R.~E., 1990. {The Strength of Weak Learnability}. Machine Learning
  5~(2), 197--227.

\bibitem[{Siwek et~al.(2009)Siwek, Osowski, and Szupiluk}]{Siwek2009}
Siwek, K., Osowski, S., Szupiluk, R., 2009. {Ensemble Neural Network Approach
  for Accurate Load Forecasting in a Power System}. International Journal of
  Applied Mathematics and Computer Science 19~(2), 303--315.

\bibitem[{Sloughter et~al.(2010)Sloughter, Gneiting, and
  Raftery}]{Sloughter2010a}
Sloughter, J.~M., Gneiting, T., Raftery, A.~E., 2010. {Probabilistic Wind Speed
  Forecasting using Ensembles and Bayesian Model Averaging}. Journal of the
  American Statistical Association 105~(489), 25--35.

\bibitem[{Soman et~al.(2010)Soman, Zareipour, Malik, and Mandal}]{Soman2010}
Soman, S.~S., Zareipour, H., Malik, O., Mandal, P., 2010. {A review of wind
  power and wind speed forecasting methods with different time horizons}. In:
  North American Power Symposium. Arlington, USA, pp. 1--8.

\bibitem[{Taylor and Buizza(2002)}]{Taylor2002}
Taylor, J., Buizza, R., 2002. {Neural network load forecasting with weather
  ensemble predictions}. IEEE Transactions on Power Systems 17~(3), 626--632.

\bibitem[{Taylor et~al.(2009)Taylor, McSharry, and Buizza}]{Taylor2009a}
Taylor, J.~W., McSharry, P.~E., Buizza, R., 2009. {Wind power density
  forecasting using ensemble predictions and time series models}. IEEE
  Transactions on Energy Conversion 24~(3), 775--782.

\bibitem[{Troldborg and S{\o}rensen(2014)}]{Troldborg2014}
Troldborg, N., S{\o}rensen, J., 2014. {A simple atmospheric boundary layer
  model applied to large eddy simulations of wind turbine wakes}. Wind Energy
  17~(April), 657--669.

\bibitem[{van Oldenborgh et~al.(2012)van Oldenborgh, Doblas-Reyes, Wouters, and
  Hazeleger}]{VanOldenborgh2012}
van Oldenborgh, G.~J., Doblas-Reyes, F.~J., Wouters, B., Hazeleger, W., 2012.
  {Decadal prediction skill in a multi-model ensemble}. Climate Dynamics
  38~(7-8), 1263--1280.

\bibitem[{Weigel et~al.(2007)Weigel, Liniger, and Appenzeller}]{Weigel2010}
Weigel, A.~P., Liniger, M.~A., Appenzeller, C., 2007. {Generalization of the
  Discrete Brier and Ranked Probability Skill Scores for Weighted Multimodel
  Ensemble Forecasts}. Monthly Weather Review 135~(7), 2778--2785.

\bibitem[{Zhang et~al.(2014)Zhang, Wang, and Wang}]{Zhang2014}
Zhang, Y., Wang, J., Wang, X., 2014. {Review on probabilistic forecasting of
  wind power generation}. Renewable and Sustainable Energy Reviews 32,
  255--270.

\bibitem[{Ziehmann(2000)}]{Ziehmann2000}
Ziehmann, C., 2000. {Comparison of a single-model EPS with a multi-model
  ensemble consisting of a few operational models}. Tellus, Series A: Dynamic
  Meteorology and Oceanography 52~(3), 280--299.

\end{thebibliography}




%
\end{document}